\documentclass[journal]{IEEEtran}

\usepackage[T1]{fontenc}
\usepackage[utf8]{inputenc}
\usepackage{cite}
\usepackage{graphicx}
\usepackage{amsmath}
\usepackage[linesnumbered, ruled]{algorithm2e}
\usepackage{amssymb}
\usepackage{standalone}
\usepackage{tikz}
\usepackage{amsfonts}
\usepackage{pgfplots}
\usepackage[numbers,sort&compress]{natbib}
\usepackage{subcaption}
\usepackage{placeins}
\usepackage{array}
\usepackage{mathrsfs}
\usepackage{algorithmic}
\usepackage{mathtools}
\usepackage{slashbox}
\usepackage{multirow}
\usepackage{listings}
\usepackage{textcomp}
\usepackage{xcolor}
\usepackage{dblfloatfix} 

\graphicspath{{figures/}}

\newcommand{\abs}[1]{\left|#1\right|}
\newcommand{\norm}[1]{\left|\left|#1\right|\right|}

\newcommand{\card}[1]{\#\{#1\}}
\newcommand{\trans}[0]{\mathsf{T}}
\newcommand{\topindex}[1]{#1}
\newcommand{\plusindex}[1]{#1}
\newcommand{\minusindex}[1]{#1}
\newcommand{\hermi}[0]{\mathsf{H}}
\newcommand{\conj}[0]{\ast}
\newcommand{\pseudoinv}[0]{\dag}
\newcommand{\bydef}[0]{\triangleq}
\newcommand{\crvec}[1]{f_{c2r}(#1)}
\newcommand{\crmat}[1]{\mathcal{F}_{c2r}(#1)}
\newcommand{\frall}[1]{\forall #1 \in \mathcal{\uppercase{#1}}}
\newcommand{\order}[1]{\mathcal{O}(#1)}
\newcommand{\lciter}[0]{p}

\begin{document}
\title{Interference Exploitation-based Hybrid Precoding with Robustness Against Phase Errors}

\author{Ganapati Hegde, ~\IEEEmembership{Student Member,~IEEE,}
	Christos Masouros,~\IEEEmembership{Senior Member,~IEEE,}
	Marius Pesavento,~\IEEEmembership{Member,~IEEE}
	\thanks{Ganapati Hegde and Marius Pesavento are with the Communication Systems Group, Technische Universit{\"a}t Darmstadt, Darmstadt 64283, Germany. (e-mail: hegde@nt.tu-darmstadt.de; pesavento@nt.tu-darmstadt.de). Christos Masouros is with the Department of Electronic \& Electrical Engineering, University College London, London WC1E7JE, U.K. (e-mail: c.masouros@ucl.ac.uk). The work of C. Masouros was supported by the Engineering and Physical Sciences Research Council Project EP/R007934/1.} 
	
	\thanks{ \textcopyright 2019 IEEE.  Personal use of this material is permitted.  Permission from IEEE must be obtained for all other uses, in any current or future media, including reprinting/republishing this material for advertising or promotional purposes, creating new collective works, for resale or redistribution to servers or lists, or reuse of any copyrighted component of this work in other works.	}
	
	\thanks{Digital Object Identifier 10.1109/TWC.2019.2917064}
}

\maketitle


\begin{abstract}
	\label{sec_abstract}
	
	Hybrid analog-digital precoding significantly reduces the hardware costs in massive MIMO transceivers when compared to fully-digital precoding at the expense of increased transmit power. In order to mitigate the above shortfall, we use the concept of constructive interference-based precoding, which has been shown to offer significant transmit power savings when compared with the conventional interference suppression-based precoding in fully-digital multiuser MIMO systems. Moreover, in order to circumvent the potential quality-of-service degradation at the users due to the hardware impairments in the transmitters, we judiciously incorporate robustness against such vulnerabilities in the precoder design. Since the undertaken constructive interference-based robust hybrid precoding problem is nonconvex with infinite constraints and thus difficult to solve optimally, we decompose the problem into two subtasks, namely, analog precoding and digital precoding. In this paper, we propose an algorithm to compute the optimal constructive interference-based robust digital precoders. Furthermore, we devise a scheme to facilitate the implementation of the proposed algorithm in a low-complexity and distributed manner. We also discuss block-level analog precoding techniques. Simulation results demonstrate the superiority of the proposed algorithm and its implementation scheme over the state-of-the-art methods.   	
\end{abstract}



\IEEEpeerreviewmaketitle


\vspace{0.25cm}

\section{Introduction}
\label{sec_introduction}

Massive multiple-input multiple-output (MIMO) system, in which the base stations (BSs) are equipped with hundreds of antennas, is one of the key pillars of the upcoming 5G mobile networks to enable immensely high spectral efficiency \cite{massive_mimo_in_hoydis, an_overview_lu}. Similar as in the conventional MIMO systems---which comprise only a few antennas (typically less than ten), the degrees of freedom resulting from the large antenna array can be used to form narrow transmit beams using precoding in a massive MIMO downlink system. The conventional precoding techniques in MIMO systems \cite{multiple_mietzner, optimal_bengtsson} are employed in the digital baseband domain, and they require a dedicated radio-frequency (RF) chain for each antenna element. However, the cost and power footprints of the RF chains pose a significant obstacle in practical implementations when the number of antennas grows large \cite{design_doan}. Therefore, devising novel precoding schemes that are appropriate in terms of hardware cost and power efficiency for massive MIMO systems is imminent.       

\begin{figure}
	\vspace{0.75cm}
	\centering
	\label{fig_system_model}
	\includegraphics[scale=1]{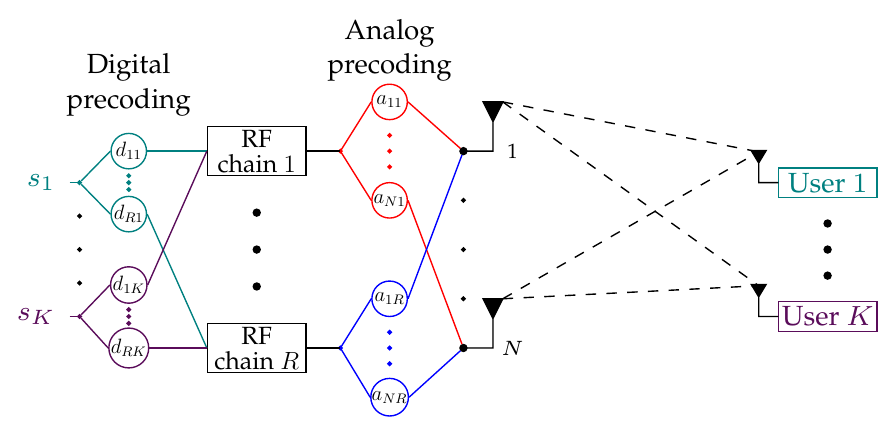}
	\vspace{0.15cm}
	\caption{Schematic diagram of hybrid analog-digital precoding system architecture.}
\end{figure}

One of the solutions to reduce the hardware costs in a massive MIMO system is hybrid analog-digital precoding \cite{low_complexity_liang, hybrid_digital_sohrabi, variable_zhang, joint_li}, which requires significantly fewer RF chains compared to the antenna elements. In this system, the RF chains are connected to the antenna elements through analog phase shifters (PSs). The hybrid precoding is performed in two stages: digital precoding in the baseband domain and analog precoding using PSs in the RF domain as shown in Fig.~\ref{fig_system_model}.

Hybrid precoding can significantly reduce the hardware costs at the expense of reduced spectral efficiency or increased transmit power to satisfy a certain quality-of-service (QoS) requirement when compared to the fully-digital precoding \cite{hybrid_molisch, hybrid_beamforming_hegde, hybrid_analog_sohrabi}. Therefore, schemes to improve energy efficiency are even more desirable in hybrid precoding-based networks than in traditional networks. The constructive interference (CI)-based precoding, in which the interference power is exploited to improve the useful signal power at the users, has shown to offer significant transmit power saving in a fully-digital multiuser downlink system \cite{exploiting_masouros, known_masouros, large_scale_amadori, rethinking_zheng, exploiting_timotheou} when compared to the conventional interference suppression or cancellation-based precoding schemes \cite{optimal_bengtsson,solution_schubert}. Therefore, it is judicious to extend the CI-based precoding to hybrid precoding architectures in order to reduce the transmit power. Some initial results in this direction are found in \cite{analog_hegde}.

Some envisioned use-cases of 5G networks, such as vehicle-to-x communication and industrial WLAN, require ultra-reliable communication \cite{towards_vannithamby}. Therefore, it is crucial to foster the precoders with robustness against interference, imperfect channel knowledge, hardware impairments, etc., in order to guarantee a certain QoS in all circumstances. In \cite{robust_wu} the authors develop a method to design the interference suppression-based hybrid precoders with robustness against multiple access interference, inter-symbol interference, and errors in the PSs. In \cite{constructive_khandaker,exploiting_masouros} the authors extend the CI-based precoding to design precoders that are robust against imperfect channel knowledge in a fully-digital precoding architecture. The CI-based precoding, in which the precoders are majorly determined by the phase values of transmit symbols and channel elements \cite{exploiting_masouros}, can be highly sensitive to the phase errors in PSs. However, to the best of our knowledge, the problem of designing the CI-based hybrid precoders with robustness against phase errors in the PSs---which is the main focus of this paper---is not considered in the literature.

In this paper, we consider symbol-level precoding\footnote{In symbol-level precoding, the precoders are updated at every symbol-interval, whereas, in block-level precoding, the precoders are kept constant for a block of symbol-intervals.} in a multiuser massive MIMO downlink system. Our goal is to design CI-based hybrid precoders that require minimum transmit power to guarantee a certain QoS to the users in the presence of phase errors in the PSs. The resulting optimization problem is nonconvex and contains infinite constraints, and thus difficult to solve optimally. To deal with the nonconvexity, we propose a method to solve the joint analog and digital precoding problem suboptimally, where we decompose the problem into two sequential subtasks, namely, analog precoding and digital precoding. In this paper, first we present methods to design the digital precoders under the premise that the analog precoders are fixed. Subsequently, we discuss various schemes to obtain the analog precoders. The main contributions of this paper are summarized below: 

$\bullet$ The CI-based robust digital precoding problem is formulated as a semi-infinite program. An iterative algorithm is proposed to solve the formulated problem. Closed-form expressions are derived to obtain the error matrices, which are required to update the constraint sets in each iteration. The convergence of the algorithm to the optimal point is proven. 

$\bullet$ A low-complexity descent-direction-based iterative scheme is devised to facilitate the distributed implementation of the proposed algorithm on parallel hardware architecture. Closed-form expressions are derived for the descent-direction and step-size, which are required in a descent-direction method \cite{nonlinear_bertsekas}.

$\bullet$ To relax the strict latency requirements on the analog precoders update, we propose block-level analog precoding. 

The paper is organized as follows. In Section~\ref{sec_system_model}, we present the system model. The optimization problem is formulated in Section~\ref{sec_problem_statement}. In Section~\ref{sec_robust_digital_precoding}, we propose an algorithm to design the optimal robust digital precoders. An iterative scheme to efficiently implement the proposed algorithm is developed in Section~\ref{sec_iterative_parallel_method}. We discuss the block-level analog precoding methods in Section~\ref{sec_analog_precoders_design}. The numerical results are presented in Section~\ref{sec_numerical_results}. Finally, Section~\ref{sec_conclusion} concludes the paper.

\textit{Notation:} Bold lower-case letters (e.g., $\mathbf{a}$) denote vectors and bold upper-case letters (e.g., $\mathbf{A}$) denote the matrices. The symbol $a_n$ represents the $n$th element of vector $\mathbf{a}$, $\mathbf{a}_r$ indicates the $r$th column of matrix $\mathbf{A}$, and $a_{nr}$ stands for the entry in the $n$th row and $r$th column of matrix $\mathbf{A}$.  $\mathcal{A}$ denotes a set, and its cardinality is indicated by $\card{\mathcal{A}}$. The letters $\mathbb{R}$ and $\mathbb{C}$ symbolize the real and complex-valued domain respectively. The operators $\mathbf{A}^{\trans}$, $\mathbf{A}^{\hermi}$, and $\mathbf{A}^{\pseudoinv}$ correspond to transpose, Hermitian, and  pseudo-inverse of matrix $\mathbf{A}$ respectively. The symbol $\odot$ represents the Hadamard (element-wise) product. The operations $\abs{\cdot}$, $\norm{\cdot}$, and $\norm{\cdot}_{\operatorname{F}}$, indicate absolute value, $\ell_2$-norm, and Frobenius-norm operations respectively. $\operatorname{exp}(\cdot)$ stands for exponential function. $\operatorname{Re}(\cdot)$ and $\operatorname{Im}(\cdot)$ denote the real part and imaginary part of a complex argument respectively. The letter $j$ stands for the imaginary unit, and $a^\conj$ indicates the complex-conjugate of complex scalar $a$.


\section{System Model}
\label{sec_system_model}

\begin{figure}[t!]
	\centering
\vspace{-0.25cm}
	\includegraphics[scale=1]{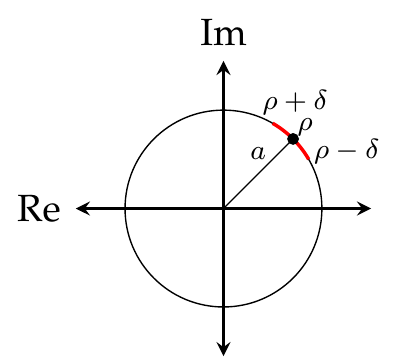}
	\caption{Phase error around the designed value of a PS.}
	\label{fig_phase_error}
\end{figure}

Consider a co-channel multiuser MIMO downlink system consisting of a BS equipped with $N$ transmit antennas and $R$ RF chains, where $R \ll N$. Let $\mathcal{K} \bydef \{1,\ldots,K\}$ denote a set of $K$ single antenna users served by the BS. The transmit symbol vector at the BS is given by $\mathbf{s} \bydef [s_1, \ldots, s_K]^{\trans}$, where the element $s_k$ indicates the symbol intended for the $k$th user. The symbols are assumed to be drawn from an \mbox{$M$-ary} phase-shift keying ($M$-PSK) constellation\footnote{Nevertheless, the proposed techniques can be extended to other modulation formats following the principles in \cite{exploiting_li}.}, and without loss of generality (w.l.o.g.) each transmit symbol is assumed to be of constant unit modulus. A digital precoder $\mathbf{d}_k \in \mathbb{C}^{R}$ is applied to the transmit symbol $s_k$ and the resulting signals are fed to the $R$ RF chains. Each RF chain is connected to all transmit antennas through analog PSs\footnote{The proposed method is equally applicable to the partially-connected hybrid architecture \cite{hybrid_analog_sohrabi}, where the values of the omitted PSs are forced to zero.}. The PSs have constant gains, and w.l.o.g. the gains are assumed to be identical, which is denoted by $a$. Let $\rho_{nr}$ denote the designed (intended) phase value of the PS that connects the $n$th antenna to the $r$th RF chain, and $a_{nr} = a\operatorname{exp}(j\rho_{nr})$ denotes the resulting complex value of the corresponding PS. Moreover, $\mathbf{a}_r \bydef [a_{1r},\ldots,a_{Nr}]^{\trans}$ forms the analog precoder applied to the output of the $r$th RF chain for $r \in \mathcal{R} \bydef \{1, \ldots, R\}$. Let $\mathbf{A} \bydef [\mathbf{a}_1, \ldots, \mathbf{a}_R]$ be the analog precoder matrix. The PSs are assumed to be imperfect, where their actual phase values can vary from their designed phase values\footnote{The phase errors are generally caused by device size limitation, inductor parasitics, loading effect, etc. The actual values depend on parameters such as frequency of operation, PS technology, and resolution, and typically vary from $2^\circ$ to $24^\circ$ \cite{sources_morton,  60_GHz_li}.}. Let $\phi_{nr}$ represent the phase error associated with a PS whose designed value is $a_{nr}$. Then the true value of the PS is given by $a\operatorname{exp}({j(\rho_{nr} + \phi_{nr})}) = a_{nr} e_{nr}$, where $e_{nr} \bydef \operatorname{exp}({j\phi_{nr}})$ represents the resulting multiplicative complex error associated with the PS. 
We assume the phase errors are bounded within a known bound $\delta$ such that $-\delta \leq \phi_{nr} \leq \delta$, as shown in Fig.~\ref{fig_phase_error}. 
Let the set $\boldsymbol{\mathcal{E}}$ denote the infinite set of all possible error matrices that are associated with the analog precoder matrix $\mathbf{A}$, i.e.,
\begin{align*}
	\boldsymbol{\mathcal{E}} \bydef \{\mathbf{E}\ \vert \ \mathbf{E} \in \mathbb{C}^{N\times R}, \abs{e_{nr}} = 1, \abs{\angle e_{nr}} \leq \delta, \frall{n}, \frall{r} \}.
\end{align*}

Let $\tilde{\mathbf{h}}_k \in \mathbb{C}^{N}$ be the frequency-flat channel vector between the BS and the $k$th user, which is assumed to be known at the BS \cite{scaling_rusek,a_coordinated_yin}. Let $n_k \sim \mathcal{CN}(0,\sigma_k^2 )$ represent the i.i.d. additive white Gaussian noise at the $k$th user. The received signal $y_{k}$ at the $k$th user can be expressed as 
\begin{equation}
y_{k} = \tilde{\mathbf{h}}_k^{\trans} (\mathbf{A}\odot\mathbf{E}) \left(  \sum_{\ell = 1}^{K} \mathbf{d}_{\ell} s_{\ell} \right) + n_{k},
\end{equation}
where the error matrix $\mathbf{E} \in \boldsymbol{\mathcal{E}}$.


\section{Problem Formulation}
\label{sec_problem_statement}

\begin{figure}[t!]
	\centering
	\includegraphics[scale=.215]{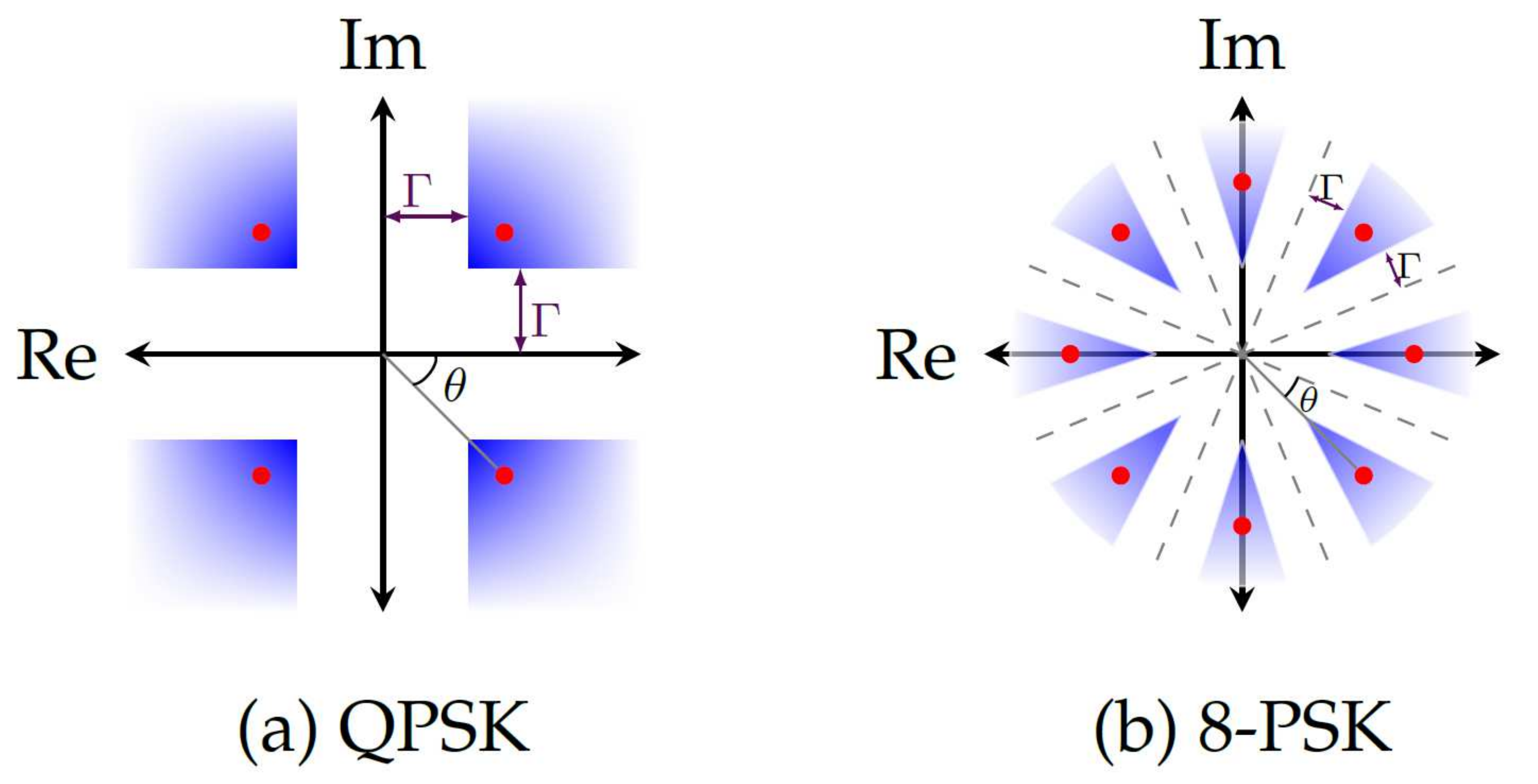}
	\caption{CI-regions (in the blue shaded area) of constellation symbols.}
	\label{fig_symbol_region}
\end{figure}

The CI-based precoding is a linear precoding technique that exploits the knowledge of the channel and data of all users to pre-equalize the transmit signals, such that the received signal at each user lies in the correct decision-region with at-least a \textit{threshold-margin} away from the corresponding decision boundaries \cite{dynamic_masouros, known_masouros, exploiting_masouros}. The part of a decision-region that is a threshold-margin away from the corresponding decision boundaries is called as \textit{constructive interference-region} (CI-region). The CI-regions of constellation symbols in the case of QPSK and 8-PSK are illustrated in Fig.~\ref{fig_symbol_region}, where $\Gamma$ represents the threshold-margin. The enforced threshold-margins control the achieved symbol-error-rates (SER) and hence the resulting QoS at the users.

In this paper, we extend the CI-based precoding concept to the hybrid precoding architecture. Our objective is to design hybrid precoders with the minimum transmit power at the BS, such that the received signal at each user lies in the CI-regions of the respective transmitted symbols. Note that, when the non-robust precoding is employed, the phase errors in the PSs can drive the received signals at the users outside the corresponding CI-regions, resulting in increased SER. To overcome this drawback, we incorporate robustness into the hybrid precoding to ensure that the received signal at each user lies in the appropriate CI-region for any error matrix $\mathbf{E} \in \boldsymbol{\mathcal{E}}$. 

\begin{figure}[t!]
	\centering
	\includegraphics[scale=.2]{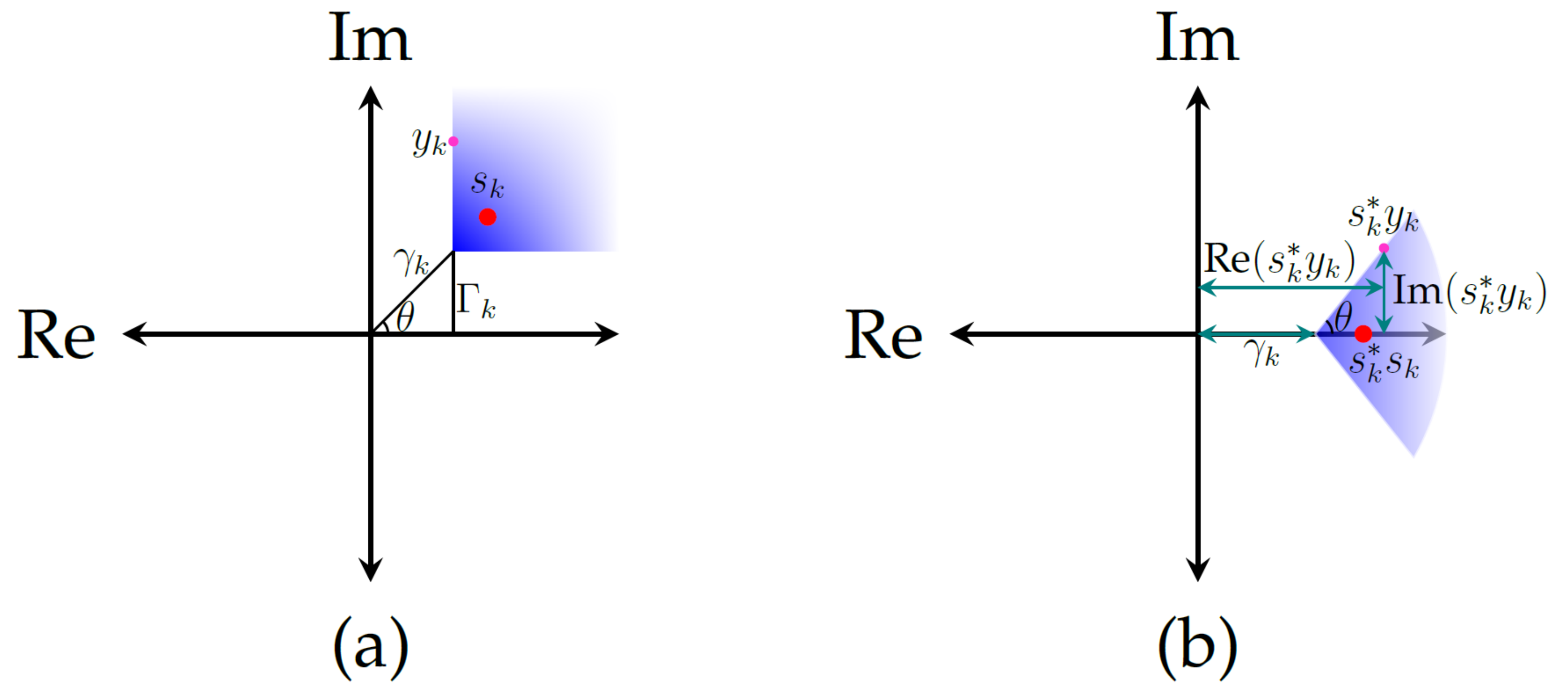}
	\caption{Rotation of transmit symbol along with the corresponding CI-region.}
	\label{fig_CI_rotation}
\end{figure}

Consider a transmit symbol $s_k$ at an angle $\theta$ and the corresponding received signal $y_k$, as shown in Fig.~\ref{fig_CI_rotation}(a). The symbol $s_k$ can be rotated onto the positive real axis by multiplying it with its conjugate $s_k^{\conj}$. Fig.~\ref{fig_CI_rotation}(b) depicts the rotated symbol $s_k^\conj s_k$ along with the corresponding CI-region and the rotated received signal $s_k^{\conj} y_k = s_{k}^{\conj}  \tilde{\mathbf{h}}_k^{\trans}   (\mathbf{A}\odot\mathbf{E}) \sum_{\ell = 1}^{K} \mathbf{d}_{\ell} s_{\ell}$. From the figure, we can deduce that the signal $s_k^{\conj} y_k$ lies in the CI-region when the following condition is fulfilled: $\frac{\abs{\operatorname{Im} \left( s_{k}^{\conj}  y_k \right) }} { \operatorname{Re} \left(s_{k}^{\conj}  y_k\right) - \gamma_k } \leq \tan \theta, \forall \mathbf{E} \in \boldsymbol{\mathcal{E}}$. 
Extending the CI-based fully-digital precoding in \cite{exploiting_masouros}, we formulate a semi-infinite program \cite{generalized_vazquez, numerical_gustafson, relaxed_wu} to implement the above-stated objective as 
\begin{subequations}
	\label{problem_grand_optimization_problem}
	\begin{align}
	\underset{\mathbf{A}, \{\mathbf{d}_{k}\}_{k \in \mathcal{K}} }{\operatorname{minimize\ }}  &\norm{ \mathbf{A} \sum_{k=1}^{K} \mathbf{d}_{k} s_{k} }^2 \label{obj_grand_optimization_problem}\\
	\operatorname{s.t.\ } &\abs{\operatorname{Im}\left( s_{k}^{\conj}  \tilde{\mathbf{h}}_k^{\trans}   (\mathbf{A}\odot\mathbf{E}) \sum_{\ell = 1}^{K} \mathbf{d}_{\ell} s_{\ell} \right)} \leq \nonumber \\ 
	&\ \ \left( \operatorname{Re} \left(s_{k}^{\conj}  \tilde{\mathbf{h}}_k^{\trans} (\mathbf{A}\odot\mathbf{E}) \sum_{\ell = 1}^{K} \mathbf{d}_{\ell} s_{\ell} \right) - \gamma_k \right) \tan \theta, \nonumber \\
	& \hspace{3.5cm}\forall \mathbf{E} \in \boldsymbol{\mathcal{E}}, \frall{k}, \label{constraint_grand_optimization_problem}\\
	&\abs{a_{nr}} = a, \quad \frall{n}, \frall{r}, \label{constraint_grand_AP_constant_magnitude}
	\end{align}
\end{subequations}
where $\theta \bydef \pi/M$, and the QoS controlling parameter $\gamma_k \bydef \Gamma_k/\sin\theta$, with $\Gamma_k$ indicating the threshold-margin at the $k$th user. In the above problem, the objective function (\ref{obj_grand_optimization_problem}) minimizes the total transmit power at the BS. The constraints in (\ref{constraint_grand_optimization_problem}) enforce the received signals to lie in the appropriate CI-regions for each user, $\forall \mathbf{E} \in \boldsymbol{\mathcal{E}}$. The constraints in (\ref{constraint_grand_AP_constant_magnitude}) enforce the constant gain of each element of the analog precoder matrix $\mathbf{A}$. By substituting  $\mathbf{h}_{k} \bydef s_{k}^{\conj} \tilde{\mathbf{h}}_k $ and treating the composite precoding term $\sum_{\ell = 1}^{K} \mathbf{d}_{\ell} s_{\ell}$ as a single precoder $\mathbf{b}$, problem (\ref{problem_grand_optimization_problem}) can be reformulated as \cite{exploiting_masouros, quality_karipidis}
\begin{subequations}
	\label{problem_virtual_multicast_problem}
	\begin{align}
	\underset{\mathbf{A},\ \mathbf{b} }{\operatorname{minimize\ }} &\norm{\mathbf{A}\mathbf{b}}^2 \label{obj_virtual_multicast_problem}     \\
	\operatorname{s.t.\ } &\abs{\operatorname{Im}\left(\mathbf{h}_{k}^{\trans} (\mathbf{A}\odot\mathbf{E}) \mathbf{b} \right) } \leq \nonumber \\ 
	&\quad \left(\operatorname{Re}\left(\mathbf{h}_{k}^{\trans} (\mathbf{A}\odot\mathbf{E}) \mathbf{b}\right)-\gamma_k\right) \tan\theta,\nonumber \\
	&\hspace{3.5cm} \forall \mathbf{E} \in \boldsymbol{\mathcal{E}}, \frall{k}, \label{constraint_virtual_multicast_problem}\\
	&\abs{a_{nr}} = a, \quad \frall{n}, \frall{r}. \label{constraint_AP_constant_magnitude}
	\end{align}
\end{subequations}
The optimal effective digital precoder $\mathbf{b}^{\star}$ of problem (\ref{problem_virtual_multicast_problem}) and the optimal original digital precoders $\mathbf{d}_{k}^{\star}$ of problem (\ref{problem_grand_optimization_problem}) are related by $\mathbf{d}_{k}^{\star} = \frac{\mathbf{b}^{\star} } {s_k K}, \frall{k}$ \cite{exploiting_masouros}.

The problem (\ref{problem_virtual_multicast_problem}) is nonconvex and difficult to solve optimally, due to the following reasons: i) bilinear coupling of analog precoder matrix $\mathbf{A}$ and the digital precoder $\mathbf{b}$, ii) the nonconvex domain of the elements of $\mathbf{A}$, and iii) the constraint in (\ref{constraint_virtual_multicast_problem}) must be satisfied $\forall \mathbf{E} \in \boldsymbol{\mathcal{E}}$, i.e., the number of constraints is infinite. We propose a sequential optimization approach that decomposes the problem into two subproblems, namely, analog precoding and robust digital precoding.  In Section~\ref{sec_robust_digital_precoding} and \ref{sec_iterative_parallel_method}, we consider the robust digital precoding and its efficient implementation under the premise that the analog precoder matrix is fixed. Subsequently, in Section~\ref{sec_analog_precoders_design} we study the analog precoder design techniques.


\vspace{0.75cm}

\section{Optimal Robust Digital Precoding}
\label{sec_robust_digital_precoding}

\begin{figure}[t!]
	\centering
	\vspace{0.5cm}
	\includegraphics[scale=0.2]{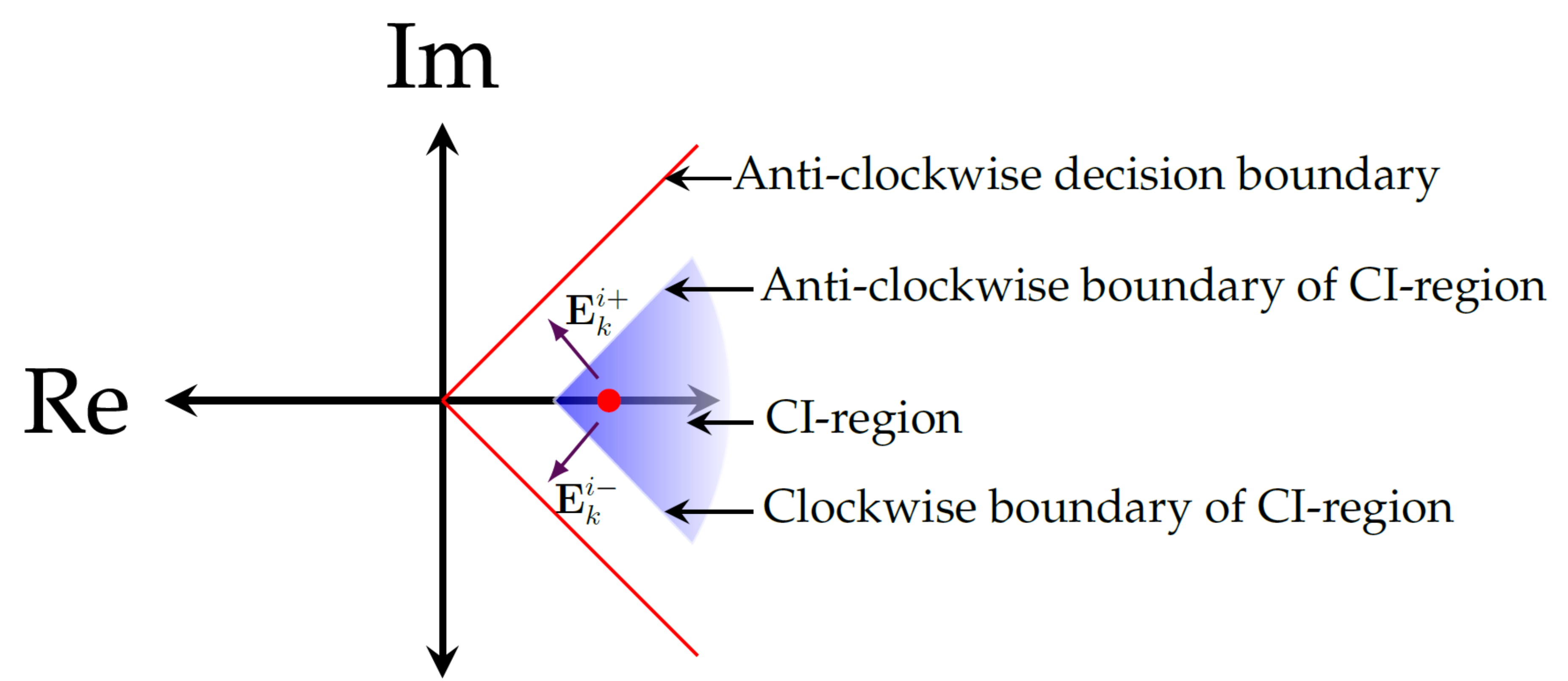}
	\vspace{0.2cm}
	\caption{Anti-clockwise and clockwise drifts of the received signals from the CI-region.}
	\label{fig_worst_error}
\end{figure}

In this section, we design the worst-case robust digital precoder $\mathbf{b}^{\star}$ of problem (\ref{problem_virtual_multicast_problem}) when the analog precoder matrix $\mathbf{A}$ is fixed to $\hat{\mathbf{A}}$. The resulting problem can be expressed as a semi-infinite problem given by
\begin{subequations}
		\label{problem_robust_virtual_multicast_problem}
		\begin{align}
		&\underset{\mathbf{b}}{\operatorname{minimize\ }} \norm{\hat{\mathbf{A}} \mathbf{b}}^2 \label{obj_robust_virtual_multicast_problem}	 \\
		&\operatorname{s.t.\ }  +{\operatorname{Im}\left(\mathbf{h}_k^{\trans} (\hat{\mathbf{A}} \odot \mathbf{E})\mathbf{b}\right) } \leq \nonumber \\
		&\ \left(\operatorname{Re}\left(\mathbf{h}_k^{\trans} (\hat{\mathbf{A}} \odot \mathbf{E}) \mathbf{b}\right) -  \gamma_k\right) \tan\theta, \ \forall \mathbf{E} \in \boldsymbol{\mathcal{E}}, \frall{k}, \label{constraint_robust_virtual_multicast_problem_1} \\
		& -{\operatorname{Im}\left(\mathbf{h}_k^{\trans} (\hat{\mathbf{A}} \odot \mathbf{E})\mathbf{b}\right) } \leq \nonumber \\
		&\ \left(\operatorname{Re}\left(\mathbf{h}_k^{\trans} (\hat{\mathbf{A}} \odot \mathbf{E}) \mathbf{b}\right) -  \gamma_k\right) \tan\theta, \  \forall \mathbf{E} \in \boldsymbol{\mathcal{E}}, \frall{k}, \label{constraint_robust_virtual_multicast_problem_2}
		\end{align}
\end{subequations}
where the constraints in (\ref{constraint_robust_virtual_multicast_problem_1}) enforce the received signal at each user to lie below the anti-clockwise boundary and the constraints in (\ref{constraint_robust_virtual_multicast_problem_2}) enforce them to lie above the clockwise boundary of the corresponding CI-region, $\forall \mathbf{E} \in \boldsymbol{\mathcal{E}}$, as shown in Fig.~\ref{fig_worst_error}. We assume that problem (\ref{problem_robust_virtual_multicast_problem}) is feasible. 
Based on the \textit{cutting plane method and alternating procedure} \cite{numerical_gustafson, relaxed_wu}, we develop an iterative algorithm to efficiently solve the formulated semi-infinite program by exploiting a structure in the problem, namely, the constant magnitude property of elements of error matrix $\mathbf{E} \in \boldsymbol{\mathcal{E}}$.

We initialize the algorithm (iteration number $i=1$) with sets $\boldsymbol{{\mathcal{E}}}^{\plusindex{i+}}_k = \{\mathbf{1}\}$ and $\boldsymbol{{\mathcal{E}}}^{\minusindex{i-}}_k = \{\mathbf{1}\}$, $\frall{k}$, where $\mathbf{1}$ is an $N$$\times$$R$ matrix with all elements equal to 1. The proposed algorithm comprises two stages in each iteration. In the first stage of the $i$th iteration we solve the following convex quadratic problem, which corresponds to the non-robust precoding problem in the first iteration.
	\begin{subequations}
		\label{problem_update_base_precoder}
		\begin{align}
		\underset{\mathbf{b}^{\topindex{i}}}{\operatorname{minimize\ }} &\norm{\hat{\mathbf{A}} \mathbf{b}^{\topindex{i}} }^2 \label{obj_update_base_precoder}	 \\
		\operatorname{s.t.\ }  &+{\operatorname{Im}\left(\mathbf{h}_k^{\trans} (\hat{\mathbf{A}} \odot \mathbf{E} ) \mathbf{b}^{\topindex{i}}\right) } \leq \nonumber \\
		&\quad \left(\operatorname{Re}\left(\mathbf{h}_k^{\trans} (\hat{\mathbf{A}} \odot \mathbf{E}) \mathbf{b}^{\topindex{i}} \right) -  \gamma_k\right) \tan\theta, \nonumber \\
		&\hspace{3cm}  \forall \mathbf{E} \in \boldsymbol{{\mathcal{E}}}^{\plusindex{i+}}_k, \frall{k}, \label{constraint_update_base_precoder_1} \\
		&-{\operatorname{Im}\left(\mathbf{h}_k^{\trans} (\hat{\mathbf{A}} \odot \mathbf{E}) \mathbf{b}^{\topindex{i}} \right) } \leq \nonumber \\
		&\quad \left(\operatorname{Re}\left(\mathbf{h}_k^{\trans} (\hat{\mathbf{A}} \odot \mathbf{E}) \mathbf{b}^{\topindex{i}} \right) -  \gamma_k\right) \tan\theta, \nonumber \\
		&\hspace{3cm}  \forall \mathbf{E} \in \boldsymbol{{\mathcal{E}}}^{\plusindex{i-}}_k, \frall{k}. \label{constraint_update_base_precoder_2}
		\end{align}
	\end{subequations}
In the subsequent iterations, this problem comprises a finite subset of constraints of problem (\ref{problem_robust_virtual_multicast_problem}): the constraint (\ref{constraint_update_base_precoder_1}) for every error matrix $\mathbf{E} \in \boldsymbol{{\mathcal{E}}}^{\plusindex{i+}}_k$; the constraint (\ref{constraint_update_base_precoder_2}) for every error matrix $\mathbf{E} \in \boldsymbol{{\mathcal{E}}}^{\plusindex{i-}}_k$, $\frall{k}$. The problem (\ref{problem_update_base_precoder}) can be solved optimally using any general purpose solver, e.g., the interior-point method. In Section~\ref{sec_iterative_parallel_method} we develop a customized scheme to solve it more efficiently. Let $\mathbf{b}^{\topindex{i} \star}$ denote the optimal solution of problem (\ref{problem_update_base_precoder}) in the $i$th iteration.     

In the second stage of the $i$th iteration, we compute the \textit{worst-case error matrices} of constraints (\ref{constraint_robust_virtual_multicast_problem_1}) and (\ref{constraint_robust_virtual_multicast_problem_2}) at $\mathbf{b} = \mathbf{b}^{\topindex{i} \star}$, $\frall{k}$. The worst-case error matrix $\mathbf{E}_{k}^{\plusindex{i+}}$ of constraint (\ref{constraint_robust_virtual_multicast_problem_1}) is defined as an error matrix $\mathbf{E} \in \boldsymbol{\mathcal{E}}$ that violates the constraint (\ref{constraint_robust_virtual_multicast_problem_1}) with the largest margin, or fulfills it with the smallest margin when the constraint is satisfied $\forall \mathbf{E} \in \boldsymbol{\mathcal{E}}$, for the $k$th user at $\mathbf{b} = \mathbf{b}^{\topindex{i} \star}$. Equivalently, the error matrix $\mathbf{E}_{k}^{\plusindex{i+}} \in  \boldsymbol{\mathcal{E}}$ causes the received signal $y_k$ at the $k$th user the farthest away from the CI-region in the anti-clockwise direction (see Fig.~\ref{fig_worst_error}), when the digital precoder is set to $\mathbf{b}^{\topindex{i} \star}$. Similarly, the worst-case error matrix of constraint (\ref{constraint_robust_virtual_multicast_problem_2}) for the $k$th user, denoted as $\mathbf{E}_{k}^{\minusindex{i-}}$, drives the received signal $y_k$ the farthest away from the corresponding CI-region in the clockwise direction. The closed-form expressions to compute  $\mathbf{E}_{k}^{\plusindex{i+}}$ and $\mathbf{E}_{k}^{\minusindex{i-}}$ are presented in a below paragraph. 
Now, if $\mathbf{E}_{k}^{\plusindex{i+}}$ violates the constraint (\ref{constraint_robust_virtual_multicast_problem_1}), then it is added to the corresponding set of error matrices, i.e., $\boldsymbol{\mathcal{E}}^{\plusindex{(i+1)+}}_k  = \boldsymbol{\mathcal{E}}^{\plusindex{i+}}_k \cup \mathbf{E}_{k}^{\plusindex{i+}}$. Similarly, if the error matrix $\mathbf{E}_{k}^{\minusindex{i-}}$ violates the constraint (\ref{constraint_robust_virtual_multicast_problem_2}), then it is included in set $\boldsymbol{\mathcal{E}}^{\minusindex{(i+1)-}}_k$, i.e., $\boldsymbol{\mathcal{E}}^{\minusindex{(i+1)-}}_k  = \boldsymbol{\mathcal{E}}^{\minusindex{i-}}_k \cup \mathbf{E}_{k}^{\minusindex{i-}}$. 
When both $\mathbf{E}_{k}^{\plusindex{i+}}$ and $\mathbf{E}_{k}^{\minusindex{i-}}$, $\frall{k}$, satisfy the constraints (\ref{constraint_robust_virtual_multicast_problem_1}) and (\ref{constraint_robust_virtual_multicast_problem_2}) respectively, we conclude that the solution of problem (\ref{problem_update_base_precoder}) is the global optimal solution of problem (\ref{problem_robust_virtual_multicast_problem}) and thus terminate the algorithm.

Optionally, in order to reduce the number of constraints of problem (\ref{problem_update_base_precoder}) in the subsequent iterations, the redundant constraints can be dropped \cite{relaxed_wu}. To this end, we identify the error matrices $\mathbf{E} \in \boldsymbol{\mathcal{E}}^{\plusindex{i+}}_k$ that result in strict inequality of the corresponding constraint in (\ref{constraint_update_base_precoder_1}) for the given digital precoder $\mathbf{b}^{\topindex{i} \star}$ and exclude them from the set $\boldsymbol{\mathcal{E}}^{\plusindex{i+}}_k$. Similarly, the error matrices $\mathbf{E} \in \boldsymbol{\mathcal{E}}^{\minusindex{i-}}_k$ that cause strict inequality of the corresponding constraint in (\ref{constraint_update_base_precoder_2}) are excluded from the set $\boldsymbol{\mathcal{E}}^{\minusindex{i-}}_k$.

\begin{figure*}[b!]
	\hrule
	\begin{align}
	&\mathbf{E}_{k}^{\plusindex{i+}} =\underset{ \abs{e_{nr}} =1, \abs{\angle e_{nr}} \leq \delta }{\operatorname{argmax\ } }  \left(+\operatorname{Im}\left(\mathbf{h}_k^{\trans} (\hat{\mathbf{A}} \odot \mathbf{E}) \mathbf{b}^{\topindex{i} \star} \right)  - \left(\operatorname{Re}\left(\mathbf{h}_k^{\trans} (\hat{\mathbf{A}} \odot \mathbf{E}) \mathbf{b}^{\topindex{i} \star} \right) -  \gamma_k\right) \tan\theta \right). \label{problem_anti-clockwise_error_matrix} \\
	&\mathbf{E}_{k}^{\minusindex{i-}} =\underset{ \abs{e_{nr}} =1, \abs{\angle e_{nr}} \leq \delta  }{\operatorname{argmax\ } }  \left(-\operatorname{Im}\left(\mathbf{h}_k^{\trans} (\hat{\mathbf{A}} \odot \mathbf{E}) \mathbf{b}^{\topindex{i} \star} \right)  - \left(\operatorname{Re}\left(\mathbf{h}_k^{\trans} (\hat{\mathbf{A}} \odot \mathbf{E}) \mathbf{b}^{\topindex{i} \star} \right) -  \gamma_k\right) \tan\theta \right). \label{problem_clockwise_error_matrix} \\
	&v_{k}^{\plusindex{i+}} = +\operatorname{Im}\left(\mathbf{h}_k^{\trans} (\hat{\mathbf{A}} \odot \mathbf{E}_{k}^{\plusindex{i+}}) \mathbf{b}^{\topindex{i} \star} \right)  - \left(\operatorname{Re}\left(\mathbf{h}_k^{\trans} (\hat{\mathbf{A}} \odot \mathbf{E}_{k}^{\plusindex{i+}}) \mathbf{b}^{\topindex{i} \star} \right) -  \gamma_k\right) \tan\theta. \label{eq_vk_plus}\\
	&v_{k}^{\minusindex{i-}} = -\operatorname{Im}\left(\mathbf{h}_k^{\trans} (\hat{\mathbf{A}} \odot \mathbf{E}_{k}^{\minusindex{i-}}) \mathbf{b}^{\topindex{i} \star} \right)  - \left(\operatorname{Re}\left(\mathbf{h}_k^{\trans} (\hat{\mathbf{A}} \odot \mathbf{E}_{k}^{\minusindex{i-}}) \mathbf{b}^{\topindex{i} \star} \right) -  \gamma_k\right) \tan\theta. \label{eq_vk_minus}
	\end{align}
\end{figure*}

\vspace{0.5cm}
\textbf{Closed-form expressions for worst-case error matrices:}
The worst-case error matrices, $\mathbf{E}_{k}^{\plusindex{i+}}$ and $\mathbf{E}_{k}^{\minusindex{i-}}$ for $k \in \mathcal{K}$, of constraints (\ref{constraint_robust_virtual_multicast_problem_1}) and (\ref{constraint_robust_virtual_multicast_problem_2}) for a given digital precoder $\mathbf{b}^{\topindex{i} \star}$ can be obtained by solving problems (\ref{problem_anti-clockwise_error_matrix}) and (\ref{problem_clockwise_error_matrix}) respectively.
These problems are nonconvex due to the nonconvex domain of optimization variables $e_{nr}, \frall{n}, \frall{r}$. We exploit the constant magnitude property of element $e_{nr}$ and derive a closed-form expression to the worst-case error matrices  (see Appendix~\ref{appendix_closed_form_solutions}), which are given by
\begin{subequations}
\begin{align}
\mathbf{E}_{k}^{\plusindex{i+}} = \mathbf{U}^{+} + j\mathbf{W}^{+}, \label{eq_anti-clockwise_error_matrix}\\
\mathbf{E}_{k}^{\minusindex{i-}} = \mathbf{U}^{-} + j\mathbf{W}^{-}, \label{eq_clockwise_error_matrix}
\end{align}
\end{subequations}
where the elements of the above matrices are computed as
\begin{subequations}
	\begin{align}
		&{u}_{nr}^{+} = \operatorname{max}\left( \cos\delta, \ \dfrac{\operatorname{Im}({z}_{nr}) \cos\theta - \operatorname{Re}({z}_{nr})\sin\theta }{\abs{{z}_{nr}}} \right), \\
		&{w}_{nr}^{+} = \dfrac{\operatorname{Re}({z}_{nr}) + \operatorname{Im}({z}_{nr}) \tan\theta}{\abs{\operatorname{Re}({z}_{nr}) + \operatorname{Im}({z}_{nr}) \tan\theta }} \sqrt{1 - ({u}_{nr}^{+})^2}, \\ 
		&{u}_{nr}^{-} = \operatorname{max}\left( \cos\delta, \ \ \dfrac{-\operatorname{Im}({z}_{nr}) \cos\theta - \operatorname{Re}({z}_{nr})\sin\theta }{\abs{{z}_{nr}}} \right), \\
		&{w}_{nr}^{-} = \dfrac{-\operatorname{Re}({z}_{nr}) + \operatorname{Im}({z}_{nr}) \tan\theta}{\abs{-\operatorname{Re}({z}_{nr}) + \operatorname{Im}({z}_{nr}) \tan\theta } } \sqrt{1 - ({u}_{nr}^{-})^2},
	\end{align}
\end{subequations}
with $\mathbf{Z} \bydef \left(\mathbf{h}_{k} {(\mathbf{b}^{\topindex{i} \star})}^{\trans}\right) \odot \hat{\mathbf{A}}$. (Indices $k$ and $i$ are dropped from the matrices $\mathbf{U}^{+}, \mathbf{W}^{+}, \mathbf{U}^{-}, \mathbf{W}^{-},$ and $\mathbf{Z}$ for notational simplicity).

Substituting the optimal solutions $\mathbf{E}_{k}^{\plusindex{i+}}$ and $\mathbf{E}_{k}^{\minusindex{i-}}$ in the objective functions of the problems (\ref{problem_anti-clockwise_error_matrix}) and (\ref{problem_clockwise_error_matrix}), we obtain the corresponding optimal values $v_{k}^{\plusindex{i+}}$ and $v_{k}^{\minusindex{i-}}$, which are given by Eq.~(\ref{eq_vk_plus}) and Eq.~(\ref{eq_vk_minus}), respectively.
A non-positive $v_{k}^{\plusindex{i+}}$ implies that $\mathbf{b}^{\topindex{i} \star}$ satisfies the constraint (\ref{constraint_robust_virtual_multicast_problem_1}) for the $k$th user $\forall \mathbf{E} \in \boldsymbol{\mathcal{E}}$. 
On the other hand, a positive value for $v_{k}^{\plusindex{i+}}$ implies that the constraint (\ref{constraint_robust_virtual_multicast_problem_1}) is violated at $\mathbf{b} = \mathbf{b}^{\topindex{i} \star}$ for the error matrix $\mathbf{E}_{k}^{\plusindex{i+}}$; a positive $v_{k}^{\minusindex{i-}}$ means the constraint (\ref{constraint_robust_virtual_multicast_problem_2}) is violated at $\mathbf{b} = \mathbf{b}^{\topindex{i} \star}$ for the error matrix
 $\mathbf{E}_{k}^{\minusindex{i-}}$, for the $k$th user.
 
The above algorithm to design the worst-case robust digital precoding is summarized in Alg.~\ref{alg_robust_algorithm}.

\textit{\textbf{Theorem 1:} When Alg.~\ref{alg_robust_algorithm} terminates after an $I$th iteration the optimal solution  $\mathbf{b}^{\topindex{I} \star}$ of problem (\ref{problem_update_base_precoder}) is equal to the optimal solution $\mathbf{b}^{\star}$ of problem (\ref{problem_robust_virtual_multicast_problem}).}    \\
Proof: see Appendix~\ref{appendix_proof_of_theorem_1}.

\textit{\textbf{Theorem 2:} The sequence, $\mathbf{b}^{\topindex{1} \star}, \mathbf{b}^{\topindex{2} \star}, \ldots,$ of optimal solutions of problem (\ref{problem_update_base_precoder}) generated by Alg.~\ref{alg_robust_algorithm} converges to the optimal solution $\mathbf{b}^{\star}$ of problem (\ref{problem_robust_virtual_multicast_problem}). }\\
Proof: see Appendix~\ref{appendix_proof_of_theorem_2}.

\begin{algorithm}
	\caption{Optimal robust digital precoding} \label{alg_robust_algorithm}
	\begin{algorithmic}[1]
		\STATE Initialize $i=1$, $\boldsymbol{\mathcal{E}}^{\plusindex{i+}}_k = \{\mathbf{1}\}$, $\boldsymbol{\mathcal{E}}^{\minusindex{i-}}_k = \{\mathbf{1}\}$, $\frall{k}$. 
		\LOOP
		\STATE Compute $\mathbf{b}^{\topindex{i} \star}$ by solving problem (\ref{problem_update_base_precoder}) [e.g., using the proposed scheme in Section~\ref{sec_iterative_parallel_method}].  
		\STATE Compute $\mathbf{E}^{\plusindex{i+}}_k$ and $\mathbf{E}^{\minusindex{i-}}_k$ using  Eq.~(\ref{eq_anti-clockwise_error_matrix}) and Eq.~(\ref{eq_clockwise_error_matrix}) respectively, $\frall{k}$. 		
		\STATE Compute $v_{k}^{\plusindex{i+}}$ and $v_{k}^{\minusindex{i-}}$ using Eq.~(\ref{eq_vk_plus}) and Eq.~(\ref{eq_vk_minus}) respectively. 		
		\STATE If $v_{k}^{\plusindex{i+}} > 0$, then add $\mathbf{E}_{k}^{\plusindex{i+}}$ to set $\boldsymbol{\mathcal{E}}^{\plusindex{i+}}_k$. Similarly, if $v_{k}^{\plusindex{i-}} > 0$, then add $\mathbf{E}_{k}^{\minusindex{i-}}$ to set $\boldsymbol{\mathcal{E}}^{\minusindex{i-}}_k$.		
		\STATE \textbf{Break}, if both $v_{k}^{\plusindex{i+}}$ and $v_{k}^{\plusindex{i-}}$ are non-positive $\frall{k}$.
		\STATE $i \leftarrow i+1$.
		\ENDLOOP
	\end{algorithmic}
\end{algorithm}


\section{Low-complexity Parallel implementation scheme}
\label{sec_iterative_parallel_method}

The major part of the computations involved in the proposed robust digital precoding algorithm is contributed from problem (\ref{problem_update_base_precoder}), which needs to be solved in every iteration. In this section, we develop a customized scheme to solve problem (\ref{problem_update_base_precoder}) in a parallelized manner in order to reap the benefits of any available parallel hardware and thus speed up the algorithm. Moreover, the proposed scheme facilitates the exploitation of special structures present in Alg.~\ref{alg_robust_algorithm}, which leads to a significantly reduced computational complexity in implementing the algorithm.    

In the following, firstly we transform the complex-valued problem (\ref{problem_update_base_precoder}) into an equivalent real-valued problem and then derive its dual problem. Subsequently, the dual problem is solved iteratively, as in \cite{aunified_yang,parallel_hegde}, to obtain the optimal solution of the primal problem by performing the following steps: first, an approximate problem is constructed for the dual problem that delivers a descent-direction of the dual problem at a given point. Next, the approximate problem is decomposed into multiple independent subproblems, which can be solved in parallel. Afterward, a closed-form expression is derived for the optimal solutions of the subproblems. Finally, we derive a closed-form expression to compute the step-size, which is required to update the current point in the descent-direction.

Let $\crmat{\mathbf{X}}$ be a function that transforms a complex matrix $\mathbf{X}$ to a real matrix $\mathbf{Y}$ such that
\begin{align}
\mathbf{Y} = \crmat{\mathbf{X}} \bydef \begin{bmatrix*}[c]
\text{Re}(\mathbf{X}), -\text{Im}(\mathbf{X}) \\
\text{Im}(\mathbf{X}), \ \ \text{Re}(\mathbf{X}) 
\end{bmatrix*}.
\end{align}
Let $\crvec{\mathbf{x}}$ be a function that transforms a complex vector $\mathbf{x}$ into a real vector $\mathbf{y}$ such that $\mathbf{y} = \crvec{\mathbf{x}} \bydef \begin{bmatrix}
\text{Re}(\mathbf{x})^{\trans}, \text{Im}(\mathbf{x})^{\trans}
\end{bmatrix}^{\trans}$. 
Let $\mathbf{M}_0 \bydef \crmat{\hat{\mathbf{A}}}$. We define the sets of the analog precoder matrices of problem (\ref{problem_update_base_precoder}) in the real-valued domain as below (the iteration index $i$ is dropped for notational convenience).
\begin{align}
\boldsymbol{\mathcal{M}}^{\plusindex{+}}_k \bydef \{\mathbf{M} = \crmat{\hat{\mathbf{A}}\odot\mathbf{E} } \ | \ \mathbf{E} \in \boldsymbol{\mathcal{E}}^{\plusindex{+}}_k \}, \quad \frall{k}, \\
\boldsymbol{\mathcal{M}}^{\minusindex{-}}_k \bydef \{\mathbf{M} = \crmat{\hat{\mathbf{A}}\odot\mathbf{E} } \ | \ \mathbf{E} \in \boldsymbol{\mathcal{E}}^{\minusindex{-}}_k \}, \quad \frall{k}.
\end{align}
Furthermore, we define the following: $\mathbf{g} \bydef \crvec{\mathbf{b}}$, $\mathbf{f}_k \bydef \crvec{\mathbf{h}_k}$, and 
\begin{align}
\mathbf{\Pi}_1 \bydef &\begin{bmatrix}
\mathbf{I},\ \ \mathbf{0}\\
\mathbf{0}, -\mathbf{I}
\end{bmatrix},\ \mathbf{\Pi}_2 \bydef \begin{bmatrix}
\mathbf{0},\ \ \mathbf{I} \\
\mathbf{I}, \ \ \mathbf{0}
\end{bmatrix},\label{eq_pi1_pi2_matrix}\\
\mathbf{p}_k \bydef\ &\mathbf{\Pi}_2\mathbf{f}_k, \quad \mathbf{q}_k \bydef \mathbf{\Pi}_1\mathbf{f}_k\tan \theta, \quad r_k \bydef \gamma_k \tan \theta.
\end{align}
In Eq.~(\ref{eq_pi1_pi2_matrix}), $\mathbf{I}$ and $\mathbf{0}$ are $N\times N$ identity and zero matrices respectively.
Now we can reformulate problem (\ref{problem_update_base_precoder}) in the real-valued domain as 
\begin{align}
\underset{\mathbf{g}}{\operatorname{minimize\ }} &\norm{\mathbf{M}_0\mathbf{g}}^2 \label{problem_base_precoder_real_domain} \\
\text{s.\ t.\ } &\left( +\mathbf{p}_k- \mathbf{q}_k\right)^{\trans} \mathbf{M} \mathbf{g} + r_k \leq 0, \ \forall \mathbf{M} \in \boldsymbol{\mathcal{M}}^{\plusindex{+}}_k,  \frall{k}, \nonumber \\
&\left(-\mathbf{p}_k - \mathbf{q}_k\right)^{\trans}\mathbf{M} \mathbf{g} + r_k \leq 0, \  \forall \mathbf{M} \in \boldsymbol{\mathcal{M}}^{\minusindex{-}}_k, \frall{k}. \nonumber
\end{align}
The Lagrangian function of the above problem can be written as
\begin{align}
\mathscr{L}(\mathbf{g}, \boldsymbol{\lambda}) &= \norm{\mathbf{M}_0 \mathbf{g}}^2 - \left(\mathbf{\Psi}\boldsymbol{\lambda} \right)^{\trans}\mathbf{g}  + \mathbf{r}^{\trans} \boldsymbol{\lambda},
\label{problem_lagrangian}
\end{align}
where $\boldsymbol{\lambda}$ denotes the vector of Lagrange multipliers. The matrix $\boldsymbol{\mathbf{\Psi}}$ and vector $\mathbf{r}$ are given by Eq.~(\ref{eq_psi}) and Eq.~(\ref{eq_r}) respectively. 
\begin{figure*}
	\hrule
	\begin{align}
	\boldsymbol{\Psi} \bydef \Big[&(\mathbf{M}_{1,1}^{\plusindex{+}})^\trans (\mathbf{q}_1 - \mathbf{p}_1),\ldots,(\mathbf{M}_{1,L^{\plusindex{+}}_1}^{\plusindex{+}})^\trans (\mathbf{q}_1 - \mathbf{p}_1),\ldots, (\mathbf{M}_{K,1}^{\plusindex{+}})^\trans (\mathbf{q}_K - \mathbf{p}_K),\ldots,(\mathbf{M}_{K,L^{\plusindex{+}}_K}^{\plusindex{+}})^\trans (\mathbf{q}_K - \mathbf{p}_K),\nonumber \\ 
	&(\mathbf{M}_{1,1}^{\minusindex{-}})^\trans (\mathbf{q}_1 + \mathbf{p}_1),\ldots,(\mathbf{M}_{1,L^{\minusindex{-}}_1}^{\minusindex{-}})^\trans (\mathbf{q}_1 + \mathbf{p}_1),\ldots, (\mathbf{M}_{K,1}^{\minusindex{-}})^\trans (\mathbf{q}_K + \mathbf{p}_K),\ldots,(\mathbf{M}_{K,L^{\minusindex{-}}_K}^{\minusindex{-}})^\trans (\mathbf{q}_K + \mathbf{p}_K)\Big], \label{eq_psi}\\
	\mathbf{r} \bydef \big[&r_1\mathbf{1}^{1\times L^{\plusindex{+}}_1},\ldots,r_K\mathbf{1}^{1\times L^{\plusindex{+}}_K},r_1\mathbf{1}^{1\times L^{\minusindex{-}}_1},\ldots,r_K\mathbf{1}^{1\times L^{\minusindex{-}}_K}  \big]^{\trans}, \label{eq_r} 	
	\end{align}%
	where $\mathbf{M}_{k,m}^{\plusindex{+}}$ is the $m$th element of $\boldsymbol{\mathcal{M}}^{\plusindex{+}}_k$, $\mathbf{M}_{k,m}^{\minusindex{-}}$ is the $m$th element of $\boldsymbol{\mathcal{M}}^{\minusindex{-}}_k$, 
	$L^{\plusindex{+}}_k \bydef \#\{\boldsymbol{\mathcal{M}}^{\plusindex{+}}_k\}$, and $L^{\minusindex{-}}_k \bydef \#\{\boldsymbol{\mathcal{M}}^{\minusindex{-}}_k\}$.
	\hrule
\end{figure*} 
Taking the infimum of the Lagrangian function $\mathscr{L}(\mathbf{g}, \boldsymbol{\lambda})$ w.r.t. $\mathbf{g}$ we obtain the dual function in terms of $\boldsymbol{\lambda}$, and subsequently, we formulate the dual problem of (\ref{problem_base_precoder_real_domain}) as 
\begin{align}
\underset{\boldsymbol{\lambda} \geq \mathbf{0}}{\operatorname{minimize\ }} &\norm{\mathbf{N}\boldsymbol{\lambda} }^2  -\mathbf{r}^{\trans} \boldsymbol{\lambda} 
\label{problem_dual_reformulated_problem}
\end{align}
where $\mathbf{N} \bydef \frac{(\mathbf{M}_0^{\pseudoinv})^{\trans} \mathbf{\Psi}}{2}$.
Note that problem (\ref{problem_base_precoder_real_domain}) is convex and it comprises only affine inequalities in $\mathbf{g}$. Therefore, according to the Slater's condition, strong duality holds for this problem when it is feasible \cite{convex_boyd}. Moreover, one of the KKT conditions dictates that the Lagrangian function (\ref{problem_lagrangian}) has a vanishing gradient w.r.t. $\mathbf{g}$ at an optimal primal point $\mathbf{g}^\star$ and an optimal dual point $\boldsymbol{\lambda}^\star$ \cite{convex_boyd}. By setting $\frac{\partial \mathscr{L}(\mathbf{g}^\star, \boldsymbol{\lambda}^\star) }{\partial \mathbf{g}} = 0$ we obtain the expression for an optimal primal point $\mathbf{g}^\star$ of problem (\ref{problem_base_precoder_real_domain})  in terms of the corresponding optimal dual point $\boldsymbol{\lambda}^\star$ as 
\begin{align}
\mathbf{g}^\star = \dfrac{(\mathbf{M}_0^{\trans} \mathbf{M}_0)^{-1}}{2} \mathbf{\Psi}\boldsymbol{\lambda}^\star.
\label{eq_optimal_real_bb}
\end{align}
In the following, we design an iterative algorithm to solve the dual problem (\ref{problem_dual_reformulated_problem}) optimally.

\textbf{\textit{Approximate problem:}} Let $W$ be the total number of elements in vector $\boldsymbol{\lambda}$ and $\mathcal{W} \bydef \{1,\ldots, W\}$. In problem (\ref{problem_dual_reformulated_problem}) the objective function is convex in each variable $\lambda_w$ for $w \in \mathcal{W}$. Therefore, based on the Jacobi theorem \cite{atutorial_palomar, aunified_yang} we construct an approximate problem for problem (\ref{problem_dual_reformulated_problem}) in the $\lciter$th iteration around a given point $\boldsymbol{\lambda}^{\topindex{\lciter}}$ as
\begin{align}
	\underset{\lambda_w\geq 0, w \in \mathcal{W}}{\operatorname{minimize\ }} &\sum_{w=1}^{W} \left( \norm{\mathbf{N}_{-w}\boldsymbol{\lambda}_{-w}^{\topindex{\lciter}} + \mathbf{n}_w \lambda_w }^2 - \mathbf{r}_{-w}^{\trans} \boldsymbol{\lambda}_{-w}^{\topindex{\lciter}} - r_w\lambda_w \right) \label{problem_approximate_problem}
\end{align}
where $\mathbf{N}_{-w}$ denotes the matrix obtained by discarding the $w$th column $\mathbf{n}_w$ from matrix $\mathbf{N}$, $\boldsymbol{\lambda}_{-w}^{\topindex{\lciter}}$ denotes the vector obtained by discarding the $w$th element from vector $\boldsymbol{\lambda}^{\topindex{\lciter}}$, and $\mathbf{r}_{-w}$ denotes the vector obtained by eliminating the $w$th element $r_w$ from vector $\mathbf{r}$.
Let $\hat{\boldsymbol{\lambda}} \bydef [\hat{\lambda}_1, \ldots, \hat{\lambda}_{W}]^{\trans}$ denote the optimal solution of this problem. Then, according to the Jacobi theorem $\hat{\boldsymbol{\lambda}}- \boldsymbol{\lambda}^{\topindex{\lciter}}$ represents a descent-direction of the objective function of problem (\ref{problem_dual_reformulated_problem}) \cite{atutorial_palomar, aunified_yang}. Therefore, the current point $\boldsymbol{\lambda}^{\topindex{\lciter}}$ can be updated in the descent-direction of the objective function of problem (\ref{problem_dual_reformulated_problem}) as 
\begin{align}
\boldsymbol{\lambda}^{\topindex{\lciter+1}} = \boldsymbol{\lambda}^{\topindex{\lciter}} + \eta^{\topindex{\lciter}}(\hat{\boldsymbol{\lambda}}  - \boldsymbol{\lambda}^{\topindex{\lciter}}), \label{eq_update_lambda}
\end{align}
where $\eta^{\topindex{\lciter}}$ is an appropriate step-size, with $0 < \eta^{\topindex{\lciter}} \leq 1$. When $\hat{\boldsymbol{\lambda}}  = \boldsymbol{\lambda}^{\topindex{\lciter}}$ the iterative algorithm has converged to the global optimal solution $\boldsymbol{\lambda}^\star$ of problem (\ref{problem_dual_reformulated_problem}).

\textbf{\textit{Decomposition of the approximate problem:}} The objective function in (\ref{problem_approximate_problem}) comprises $W$ summands, where each summand contains only one optimization variable $\lambda_w$. Therefore, we can decompose problem (\ref{problem_approximate_problem}) into $W$ independent subproblems \cite{aunified_yang}, each involving only one optimization variable $\lambda_w$, as
\begin{align}
	\hat{\lambda}_w = \underset{\lambda_w \geq 0}{\operatorname{argmin\ }} & \norm{\mathbf{N}_{-w}\boldsymbol{\lambda}_{-w}^{\topindex{\lciter}} + \mathbf{n}_w \lambda_w }^2 - r_w\lambda_w, \label{problem_subproblem}
\end{align}
$\frall{w}$, where the constant $\mathbf{r}_{-w}^{\trans} \boldsymbol{\lambda}_{-w}^{\topindex{\lciter}}$ has been dropped without affecting the optimal solution.

\textbf{\textit{Closed-form solution of the subproblems:}} 
The objective function in subproblem (\ref{problem_subproblem}) is convex in $\lambda_w$, and it comprises only an affine inequality, namely, $\lambda_w \geq 0$. According to Slater's condition the strong duality holds for the subproblem and its dual, and KKT conditions are satisfied by the primal and dual optimal points \cite{convex_boyd}. The Lagrangian of subproblem (\ref{problem_subproblem}) can be written as
\begin{align}
\mathscr{L}(\lambda_w, \mu_w) = \norm{\mathbf{N}_{-w}\boldsymbol{\lambda}_{-w}^{\topindex{\lciter}} + \mathbf{n}_w \lambda_w }^2 - r_w\lambda_w -\mu_w\lambda_w,
\end{align}
where $\mu_w$ is the Lagrange multiplier.  
Using the KKT conditions we derive a closed-form expression for $\hat{\lambda}_w$ as
\begin{align}
	\hat{\lambda}_w = \operatorname{max}\left(0, \dfrac{1}{\norm{\mathbf{n}_w}^2}\left( \dfrac{ r_w}{2} - \mathbf{n}_w^{\trans} \mathbf{N}_{-w}\boldsymbol{\lambda}_{-w}^{\topindex{\lciter}} \right) \right). \label{eq_closed_form_expression_lambda_hat}
\end{align}

\textbf{\textit{Optimal step-size computation:}} Based on the exact line search method \cite{aunified_yang}, we can formulate an optimization problem to compute the optimal step-size $\eta^{\topindex{\lciter}}$ that minimizes the objective function of problem (\ref{problem_dual_reformulated_problem}) between the current point $\boldsymbol{\lambda}^{\topindex{\lciter}} $ and the descent-direction $\hat{\boldsymbol{\lambda}}$ as 
\begin{align}
\eta^{\topindex{\lciter}} = \underset{0 \leq \eta \leq 1}{\operatorname{argmin\ }} \underbrace{\norm{\mathbf{N} \left(\boldsymbol{\lambda}^{\topindex{\lciter}} + \eta(\hat{\boldsymbol{\lambda}}  - \boldsymbol{\lambda}^{\topindex{\lciter}})\right) }^2 - \mathbf{r}^{\trans} \left( \boldsymbol{\lambda}^{\topindex{\lciter}} + \eta(\hat{\boldsymbol{\lambda}}  - \boldsymbol{\lambda}^{\topindex{\lciter}}) \right)}_{\mathring{f}(\eta)}. \label{problem_step_size_optimization}
\end{align}
The function $\mathring{f}(\eta)$ in the above problem is convex and differentiable in $\eta$. Differentiating $\mathring{f}(\eta)$ w.r.t. $\eta$ and equating the gradient to zero, we obtain a closed-form expression for the optimal solution $\eta^{\topindex{\lciter}}$ of problem (\ref{problem_step_size_optimization}) as
\begin{align}
\eta^{\topindex{\lciter}} = \left[ \dfrac{-2(\mathbf{N}\boldsymbol{\lambda}^{\topindex{\lciter}})^{\trans}\mathbf{N}(\hat{\boldsymbol{\lambda}} - \boldsymbol{\lambda}^{\topindex{\lciter}}) + \mathbf{r}^{\trans}(\hat{\boldsymbol{\lambda}} - \boldsymbol{\lambda}^{\topindex{\lciter}}) }{ 2\left(\mathbf{N}( \hat{\boldsymbol{\lambda}} - \boldsymbol{\lambda}^{\topindex{\lciter}} ) \right)^{\trans} \mathbf{N} (\hat{\boldsymbol{\lambda}} - \boldsymbol{\lambda}^{\topindex{\lciter}})  }\right]_0^1. \label{eq_step_size}
\end{align}

\textbf{\textit{Termination:}} When $\hat{\boldsymbol{\lambda}} = \boldsymbol{\lambda}^{\topindex{\lciter}}$ the iterative algorithm has converged to the global optimal solution of problem (\ref{problem_dual_reformulated_problem}) \cite{aunified_yang}. If a finite numerical precision is sufficient, the iterations can be terminated when $||\boldsymbol{\lambda}^{(\lciter+1)} - \boldsymbol{\lambda}^{(\lciter)}|| \leq \varepsilon$, where $\varepsilon$ is a sufficiently small positive scalar.

The above-proposed scheme to solve problem (\ref{problem_update_base_precoder}) is summarized in Alg.~\ref{alg_iterative_parallel_method}. During the implementation of the scheme, we exploit structures in Alg.~\ref{alg_robust_algorithm} and Alg.~\ref{alg_iterative_parallel_method} to reduce the computational complexity. We note that in matrices $\boldsymbol{\Psi}$ and $\mathbf{N}$ in the proposed scheme, each column corresponds to a constraint of problem (\ref{problem_update_base_precoder}). By exploiting the fact that problem (\ref{problem_update_base_precoder}) retains all constraints it obtained in the previous iterations, we can reuse the corresponding columns of $\boldsymbol{\Psi}$ and $\mathbf{N}$ and compute only those columns that correspond to the newly added constraints\footnote{It is observed that, as the algorithm progresses the number of constraints inherited from the previous iterations is significantly larger than the number of newly added constraints.}. Similarly, we can reuse the computations $\norm{\mathbf{n}_w}^{-2}$ and $\mathbf{n}_w^{\trans} \mathbf{N}_{-w}$ that correspond to the inherited constraints in Eq.~(\ref{eq_closed_form_expression_lambda_hat}). Furthermore, Eq.~(\ref{eq_update_lambda}) can be exploited to reduce the computational complexity associated with computing the step-size $\eta^p$ in Eq.~(\ref{eq_step_size}) by reusing the previously computed terms $\mathbf{N}\boldsymbol{\lambda}^{\topindex{\lciter-1}}$,  $\mathbf{N}\boldsymbol{\hat{\lambda}}^{\topindex{\lciter-1}}$,$\mathbf{r}^{\trans}\boldsymbol{\lambda}^{\topindex{\lciter-1}}$, and $\mathbf{r}^{\trans}\boldsymbol{\hat{\lambda}}^{\topindex{\lciter-1}}$. We also exploit the fact that matrices $\hat{\mathbf{A}}$ and $\mathbf{M}_0$ remain the same in all iterations of Alg.~\ref{alg_robust_algorithm}. Therefore, the matrix $ \boldsymbol{\Delta} \bydef \frac{(\mathbf{M}_0^{\trans} \mathbf{M}_0)^{-1}}{2}$ in Eq.~(\ref{eq_optimal_real_bb}), and $\frac{(\mathbf{M}_0^{\pseudoinv})^{\trans}}{2} = \mathbf{M}^{\conj}\boldsymbol{\Delta}$ in computing $\mathbf{N}$ can be reused in every iteration. 
\begin{algorithm}
	\caption{Low-complexity parallel implementation scheme} 
	\label{alg_iterative_parallel_method}
	\begin{algorithmic}[1]
		\STATE Initialize $\lciter=1$ and $\boldsymbol{\lambda}^{(1)}$ to any non-negative values.
		\LOOP
		\STATE Compute the descent-direction $\hat{\boldsymbol{\lambda}}$ using Eq.~(\ref{eq_closed_form_expression_lambda_hat}) \quad [each element in $\hat{\boldsymbol{\lambda}}$ can be computed in parallel].
		\STATE Compute the step-size $\eta^{\topindex{\lciter}}$ using Eq.~(\ref{eq_step_size}).
		\STATE Update the current point to $\boldsymbol{\lambda}^{\topindex{\lciter+1}}$ using Eq.~(\ref{eq_update_lambda}).
		\STATE\textbf{Break}, if $||\boldsymbol{\lambda}^{(\lciter+1)} - \boldsymbol{\lambda}^{(\lciter)}|| \leq \varepsilon$.
		\STATE $\lciter\leftarrow \lciter+1$.
		\ENDLOOP
		\STATE Compute $\mathbf{g}^\star$ from $\boldsymbol{\lambda}^\star$ using Eq.~(\ref{eq_optimal_real_bb}). Subsequently, obtain complex vector $\mathbf{b}^\star$ from $\mathbf{g}^\star$.
	\end{algorithmic}
\end{algorithm}


\section{Block-level Analog precoding}
\label{sec_analog_precoders_design}

In this section, we discuss techniques to design analog precoders in the CI-based hybrid precoding setting. Two types of analog precoders are generally used in hybrid precoding systems, namely, continuous-valued analog precoders \cite{hybrid_digital_sohrabi} and codebook-based analog precoders \cite{limited_alkhateeb,hybrid_molisch}. A continuous-valued analog precoder has more degrees of freedom when compared to its counterpart, as each of its elements can take any phase value between 0 and $2\pi$. However, its realization requires expensive high-resolution tunable PSs. In contrast, in codebook-based analog precoding, the analog precoders are selected from a predefined codebook that is commonly realized in hardware with switchable spatial filter banks composed of inexpensive fixed PSs \cite{joint_hegde}. Due to a lesser degrees of freedom, the codebook-based analog precoders require an increased transmit power to fulfill a certain QoS as compared to the continuous-valued analog precoders.

Paper \cite{analog_hegde} compares the performance of different symbol-level analog precoder design techniques in a CI-based hybrid precoding system. Employing the symbol-level analog precoders, however, can become inappropriate in many scenarios, such as ultra-low latency applications of 5G networks having symbol duration requirement of few microseconds \cite{towards_vannithamby}. 
In such cases, the symbol-level analog precoding can cause drastic performance degradation in hybrid precoding systems with inexpensive PSs having the transient response time in the order of microseconds (e.g., PSs comprising RF MEMS \cite{design_bansal}). 
To overcome this shortcoming, we propose the block-level analog precoding, where an analog precoder matrix that is suitable for a block of $T$ symbol-intervals is designed. 
We choose $T \leq T_{\text{c}}$, where $T_{\text{c}}$ is the coherence time of the channel so that the block-level analog precoder matrix can be designed using the known constant channel matrix. In the following, we extend the methods of \cite{analog_hegde} to the block-level analog precoding.


\subsection{Continuous-valued analog precoder design}


\subsubsection{Conjugate phase of channel (CPC) method} In this method, the BS assigns an RF chain to each user. Then, the array gain between the $k$th user and the associated RF chain is maximized by assigning the conjugate phase values of the elements of the channel vector $\tilde{\mathbf{h}}_k$ to the corresponding elements of the analog precoder $\mathbf{a}_k$, i.e., \mbox{$a_{nk} = a \exp({-j\beta_{nk}})$}, where $\beta_{nk}$ indicates the phase value of the $n$th element of channel vector $\tilde{\mathbf{h}}_k$ \cite{low_complexity_liang}. We remark that in this method the analog precoder matrix is independent of transmit symbol vector and remain the same as long as the channel is constant. Thus, the method is inherently suitable for block-level analog precoder design.     


\subsection{Codebook-based analog precoder design}
In the codebook-based analog precoder design techniques, the analog precoders are chosen from a predefined set $\mathcal{C} \bydef \{\mathbf{c}_1, \ldots, \mathbf{c}_C\}$, where $C \geq R$. Let $\mathbf{C} \bydef [\mathbf{c}_1, \ldots, \mathbf{c}_C]$ be the corresponding codebook matrix.     


\subsubsection{Margin widening and selection operator (MWASO)}
\label{subsubsec_mwaso}

In \cite{analog_hegde} a sparsity-based analog precoder selection technique, termed as MWASO, is devised to select $R$ analog precoders from the codebook that maximize a utility function. Here, we extend this technique to enable block-level analog precoding over $T$ symbol-intervals by formulating a block-sparsity-based convex optimization problem \cite{block_elhamifar, block_steffens} as
\begin{subequations}
\label{problem_group_lasso}
\begin{align}
	&\underset{\Upsilon \in \mathbb{R}, \{{\mathbf{x}}_{t}\}_{t \in \mathcal{T}} }{\operatorname{minimize\ }} \Upsilon + \epsilon \norm{ {\mathbf{X}}}_{2,1} \label{obj_group_lasso} \\
	&\operatorname{s.t.\ } \abs{\operatorname{Im}\left(s_{k}^{\conj}  \tilde{\mathbf{h}}_k^{\trans}  \mathbf{C} {\mathbf{x}}_{t} \right) } \leq \left(\operatorname{Re}\left(s_{k}^{\conj}  \tilde{\mathbf{h}}_k^{\trans}  \mathbf{C} {\mathbf{x}}_{t} \right) -  \left( \gamma_k - \Upsilon \right) \right) \tan\theta, \nonumber \\
	&\hspace{5cm} \frall{k}, \frall{t}. \label{constraint_group_lasso}
\end{align}
\end{subequations}
In this problem, $\gamma_k - \Upsilon$ determines the minimum margin between the received signals and the decision boundaries of the associated transmit symbols of the $k$th user over all $T$ symbol-intervals. The optimization variable $\Upsilon$ in the objective function along with the constraints in (\ref{constraint_group_lasso}) forces the received signals towards the interior of the CI-region for all $K$ users over all $T$ symbol-intervals. The optimization matrix ${\mathbf{X}} \bydef [{\mathbf{x}}_{1}, \ldots, {\mathbf{x}}_{T}] \in \mathbb{C}^{C \times T}$ acts as the selection operator. The mixed $\ell_{2,1}$ norm in the objective function promotes row sparsity on matrix ${\mathbf{X}}$ \cite{block_steffens}, thereby allowing the selection of analog precoders from codebook matrix $\mathbf{C}$ that are appropriate for all $T$ symbol-intervals. The positive scalar $\epsilon$ is an appropriate weighting factor, which can be chosen, e.g., using bisection method, to force the number of non-zero rows in ${\mathbf{X}}$ to $R$. Subsequently, the columns of the codebook matrix $\mathbf{C}$ that correspond to the non-zero rows of the optimal solution ${\mathbf{X}}^\star$ form the analog precoder matrix $\hat{\mathbf{A}}$. We remark that the digital precoders $\mathbf{x}_t,  \frall{t}$ are not optimal due to the row-sparse promoting term $\norm{ {\mathbf{X}}}_{2,1}$ in the objective function. Therefore, they are not reused while computing the robust digital precoders in Section~\ref{sec_robust_digital_precoding}. We also note that this method is suitable for a fully-connected hybrid precoding architecture with all PSs having an identical gain. 


\subsubsection{Best matching code selection (BMCS) method}
In this method, for each user the analog precoder from the codebook $\mathcal{C}$ that maximizes the inner product with its channel vector is selected \cite{analog_hegde}. Similar to the CPC method, this method designs the analog precoders  independent of transmit symbol vectors, and hence it is inherently suitable for block-level analog precoding.


Remark: The proposed CI-based robust hybrid precoding assumes the conventional hybrid precoding architecture typically considered in the literature. Therefore, the circuitry power consumption of the conventional
hybrid precoding (e.g., detailed in \cite{hybrid_mimo_rial}) and that of the proposed precoding are the same when we employ the proposed CPC and BMCS methods. When the MWASO method is employed at every symbol, even though it needs frequent switching of codes using RF switches, due to significant saving of transmit power compared to the conventional precoding (up to a few watts as demonstrated in the next section) the operational power associated with the switches (few milliwatts \cite{hybrid_mimo_rial}) becomes negligibly small.


\section{Numerical Results}
\label{sec_numerical_results}

For the simulation, we employ the geometric channel model \cite{spatially_ayach,low_complexity_liang}, which is given by $\tilde{\mathbf{h}}_k = \sqrt{\frac{N}{L}} \sum_{\ell =1}^{L} \alpha_\ell^k \mathbf{u}(\Phi_\ell^k)$, where $L$ denotes the number of propagation paths ($L$ is set to 15 in the simulation),  $\alpha_\ell^k \sim \mathcal{CN}(0,1)$ is the complex gain of the $\ell$th path, $\mathbf{u}(\Phi_\ell^k)$ denotes the uniform linear array (ULA) response vector in the azimuth angle $\Phi_\ell^k$. The angle $\Phi_\ell^k$ is drawn from the uniform distribution over [0, $2\pi$]. The ULA response vector is given by $\mathbf{u}(\Phi) = \frac{1}{\sqrt{N}}[1, \exp(j\bar{k}d\sin(\Phi)), \ldots, \exp(j(N-1)\bar{k}d\sin(\Phi))]^\trans$, where $\bar{k} = 2\pi/\lambda$ and the inter-element spacing $d$ is set to half-wavelength $\lambda/2$. The phase errors are distributed uniformly on the interval $[-\delta, +\delta]$. 

In interference suppression-based precoding systems, the SINR metric is generally used to measure the quality of received signals, as the SINR controls the achieved SER. However, in CI-based precoding systems the interference plays a constructive role, and it does not necessarily cause symbol-errors; therefore, the SINR is not an appropriate metric to measure the quality of received signals in this system. In order to quantify the received signal quality in CI-based precoding in a noisy environment, we introduce a metric called \textit{Threshold-margin-to-Noise power Ratio} (TNR), which is defined as $\text{TNR}_k \bydef \frac{\Gamma_k}{\sigma_k^2}$. It is the ratio of the margin between the CI-region and the corresponding decision boundaries to the noise power, and directly influences the achieved SER.
The empirical relations between SNR, TNR, and SER for different modulation schemes are provided in Appendix~\ref{appendix_ser_snr_tnr_relation}.


\subsection{QoS degradation due to errors in PSs}
\label{subsec_qos_degradation}
\begin{figure}[h!]
	\centering
	\includegraphics[scale=0.8]{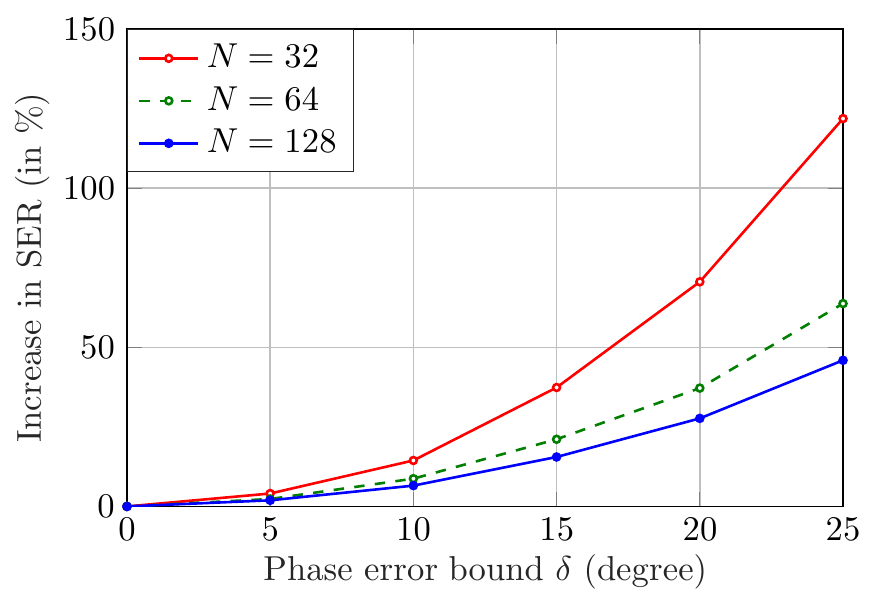}
	\caption{SER increase vs. phase error bound $\delta$ in the case of CI-based non-robust hybrid precoding for $R=K=M=4$, TNR = 2, $T=1$, and CPC analog precoder design.}
	\label{fig_ser_vs_delta_nonrobust}
\end{figure}

Fig.~\ref{fig_ser_vs_delta_nonrobust} plots the percentage increase in SER for different phase error bound $\delta$, when the non-robust CI-based hybrid precoding is employed. In the figure, we notice that as the value of $\delta$ increases the SER increases significantly. When the number of transmit antennas is relatively small, the increase in SER is substantial (approx. 125\% for $N=32$). On the other hand, the proposed robust precoding is designed to handle the worst-case scenario; thus it completely eliminates the symbol-errors resulting from phase errors in PSs (i.e., 0\% SER increase).   


\subsection{Proposed vs. conventional robust hybrid precoding}
\label{subsection_conventional_robust_precoding}

A conventional approach to obtaining the robust digital precoders is to design non-robust digital precoders targeting a larger QoS than the required QoS. This technique provides robustness against errors by assigning an extra power to the digital precoders, when compared to the power required to achieve the actual QoS in the error-free scenario \cite{worst_case_wang, a_robust_pascual}. This method can be extended to the CI-based hybrid precoding by appropriately choosing a new TNR value for the non-robust precoding that achieves a similar SER performance in the presence of phase errors in the PSs and additive noise at the users, as that of the optimal worst-case robust digital precoders. In our simulation, we choose the new TNR values using the empirical relation between $\delta$, TNR, and SER given in Table~\ref{table_delta_tnr_ser_relation}.

\begin{table}[h!]
	\setlength\tabcolsep{3.5pt}
	\caption{Performance comparison of the proposed optimal and the conventional CI-based robust hybrid precoding for $N=128$, $R=K=M=4$, $T=1$, and CPC analog precoder design.}
	\begin{center}
		\begin{tabular}{| c | c | c | c | c | c | c  | } \hline
			\multirow{2}{*}{$\ \ \delta$} & \multicolumn{2}{c|}{Optimal robust} & \multirow{2}{*}{SER} & \multicolumn{2}{c|}{Conventional robust} & \multirow{2}{*}{ $\frac{(P_{\text{conv}} - P_{\text{opt}})}{P_{\text{opt}}} $} \\ \cline{2-3} \cline{5-6} 
			&TNR & $P_{\text{opt}}$ (watt) & & TNR & $P_{\text{conv}}$ (watt) & \\ \hline \hline
			1\textdegree   & 2   & 0.4133  &  2.5e-3 &   2.2603 &   0.4631 &   12.0\% \\ \hline
			2\textdegree   & 2   & 0.4816  &  1.1e-3 &   2.4195 &   0.5280 &   9.6\% \\ \hline
			3\textdegree   & 2   & 0.5695  &  4.0e-4 &   2.4989 &   0.5834 &   2.4\% \\ \hline
			4\textdegree   & 2   & 0.6855  &  1.0e-4 &   2.8731 &   0.7376 &   7.6\% \\ \hline          
		\end{tabular}
	\end{center}
	\label{table_tnr_tuning}
\end{table}

In Table~\ref{table_tnr_tuning}, we design robust hybrid precoders for different values of $\delta$ using the proposed algorithm to achieve a TNR~=~2, and compute the required transmit power $P_{\text{opt}}$ and the resulting SER. Then we compute the TNR value required to achieve a similar SER performance for the given $\delta$ with non-robust precoding using Table~\ref{table_delta_tnr_ser_relation}. For this tuned TNR value we design the non-robust digital precoders and compute the resulting transmit power $P_{\text{conv}}$. The table reveals that the conventional method requires significantly more transmit power when compared to the optimal robust precoding method. Since the optimal method guarantees the worst-case robustness, and the conventional method ensures only a statistical SER performance, the difference between transmit powers of the two methods tends to decrease as $\delta$ increases.

\begin{table}[h!]
	\caption{SER achieved by the CI-based non-robust precoding for a range of phase error bound $\delta$ and TNR for $N=128$, $R=K=4$, $T=1$, QPSK modulation, and CPC method.}
	\begin{center}
		\begin{tabular}{| c | c | c | c | c |} \hline
			$\delta$ (deg) & TNR =  2.0 & TNR = 2.5 & TNR = 3.0 \\ \hline \hline
			0 & 4.665 x$10^{-3}$ & 4.120 x$10^{-4}$ & 2.320 x$10^{-5}$ \\ \hline 
			1 & 4.667 x$10^{-3}$ & 4.126 x$10^{-4}$ & 2.333 x$10^{-5}$ \\ \hline 
			2 & 4.680 x$10^{-3}$ & 4.138 x$10^{-4}$ & 2.340 x$10^{-5}$ \\ \hline 
			3 & 4.700 x$10^{-3}$ & 4.148 x$10^{-4}$ & 2.351 x$10^{-5}$ \\ \hline 
			4 & 4.739 x$10^{-3}$ & 4.185 x$10^{-4}$ & 2.355 x$10^{-5}$ \\ \hline 
			5 & 4.755 x$10^{-3}$ & 4.208 x$10^{-4}$ & 2.370 x$10^{-5}$ \\ \hline 
			6 & 4.776 x$10^{-3}$ & 4.248 x$10^{-4}$ & 2.400 x$10^{-5}$ \\ \hline 
			7 & 4.847 x$10^{-3}$ & 4.290 x$10^{-4}$ & 2.460 x$10^{-5}$ \\ \hline 
			8 & 4.848 x$10^{-3}$ & 4.373 x$10^{-4}$ & 2.490 x$10^{-5}$ \\ \hline 
			9 & 4.929 x$10^{-3}$ & 4.431 x$10^{-4}$ & 2.533 x$10^{-5}$ \\ \hline 
			10 & 4.970 x$10^{-3}$ & 4.555 x$10^{-4}$ & 2.600 x$10^{-5}$ \\ \hline 
		\end{tabular}
	\end{center}
	\label{table_delta_tnr_ser_relation}
\end{table}


\subsection{CI-based precoding vs. state-of-the-art precoding schemes}
\label{subsection_state_of_the_art}

In this subsection, we compare the SER achieved by the proposed CI-based hybrid precoding employing the CPC method (CI-HP) with that of the following state-of-the-art hybrid precoding schemes: the PZF method proposed in \cite{low_complexity_liang},  interference suppression-based hybrid precoding method (IS-HP) proposed in \cite{hybrid_digital_sohrabi}. We also include in the figure, the performance of CI-based fully-digital precoding  (CI fully-DP) and conventional fully-digital precoding (Conv. fully-DP) \cite{solution_schubert} for reference. 

In the CI-based precoding problems (both hybrid and fully-digital precoding), the objective is to minimize the transmit power for a given TNR (accordingly a fixed SER) and fixed $M$ (hence fixed data rate). However, the considered competing methods aim to maximize the data rate (or SINR) for a given power budget. In order to facilitate a fair comparison, firstly we compute the power required by the CI-based methods to achieve a chosen TNR for a fixed modulation order $M$. Subsequently, the resulting powers are used as power budgets in the competing methods to compute the precoders and the corresponding SNRs. Moreover, we utilize the empirical relation between SNR, TNR, and SER given in Appendix~\ref{appendix_ser_snr_tnr_relation}, and obtain \textit{SER vs. transmit power} relations for all methods. We use $R=5$ RF chains for IS-HP method (this method requires $R > K$) and $R=4$ for the remaining hybrid precoding methods.  
\begin{figure}[h!]
	\centering
	\includegraphics[scale=0.71]{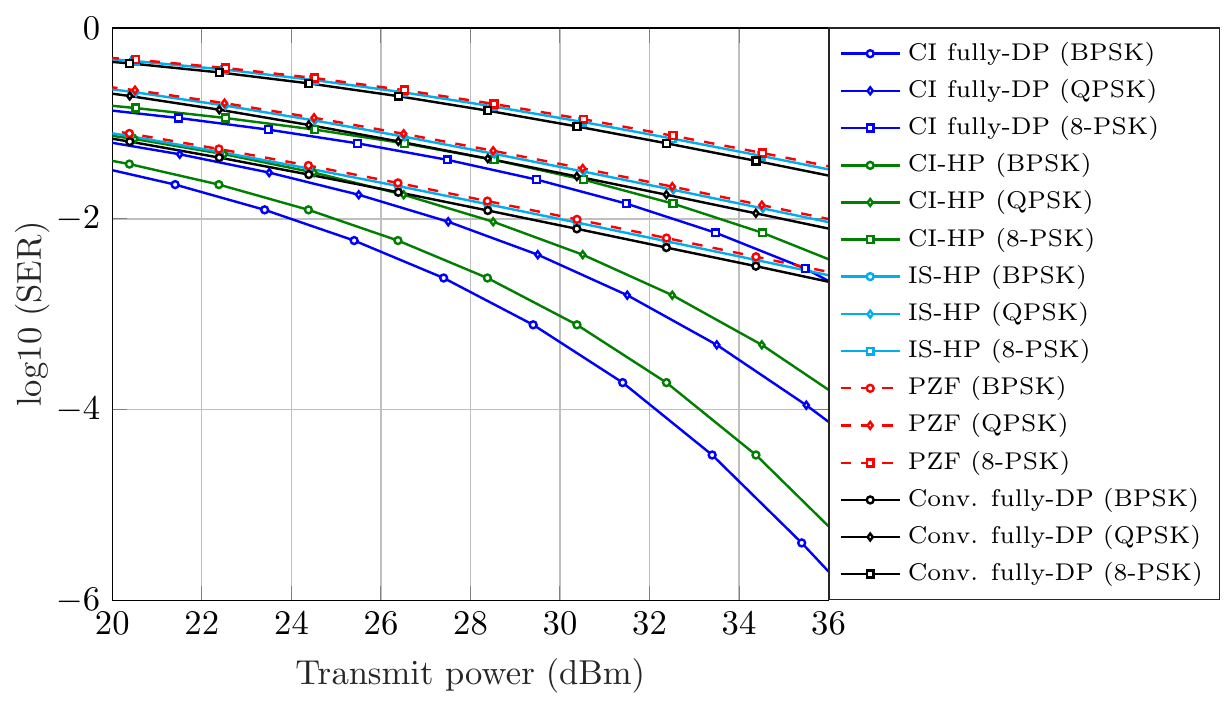}
	\caption{SER comparison of the proposed hybrid precoding, the state-of-the-art hybrid precoding, and fully-digital precoding methods for $N=128$, $K=4$, $\mathbf{E} = \mathbf{0}$, and $T=1$.}
	\label{fig_serconv_serci_vs_power_N_128_K_4_R_4_ctype_2}    
\end{figure}

In Fig.~\ref{fig_serconv_serci_vs_power_N_128_K_4_R_4_ctype_2} we notice that the proposed method considerably reduces the SER when compared to the competing hybrid and the conventional fully-digital precoding methods for all considered modulation schemes (approx. 500x for BPSK with transmit power = 36 dBm). The figure reveals that the proposed method saves a significant amount of transmit power (up to a few watts) to achieve a given SER when compared to the competing methods.     
     

\subsection{CI-based hybrid precoding vs. fully-digital precoding}
\label{subsec_ci_vs_fully_dp}

\begin{figure}[t!]
	\centering
	\includegraphics[scale=0.9]{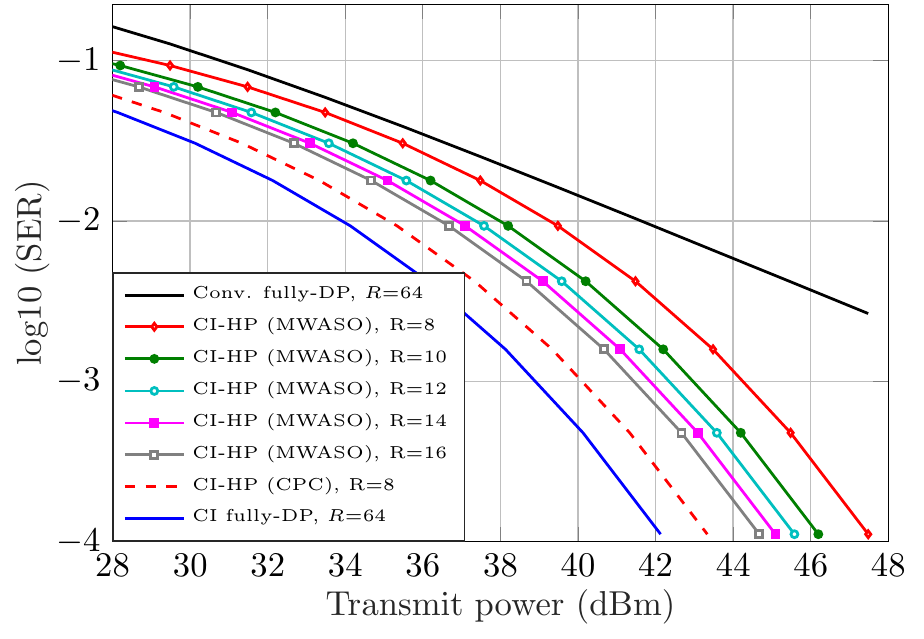}
	\caption{Performance comparison of the proposed CI-based hybrid precoding (CI-HP), CI-based and conv. fully-digital precoding for $N=64$, $K=8$, $M=4$, $\mathbf{E} = \mathbf{0}$, and $T=1$.}
	\label{fig_ser_hbf_vs_fdbf_N_64_K_8_ctype_2}
\end{figure}

Fig.~\ref{fig_ser_hbf_vs_fdbf_N_64_K_8_ctype_2} compares the SER achieved by the proposed CI-based hybrid precoding (CI-HP) with that of the optimal CI-based fully-digital precoding \cite{exploiting_masouros} (CI fully-DP) and conventional fully-digital precoding \cite{solution_schubert,optimal_bengtsson} (Conv. fully-DP). Both CI-based and conventional fully-digital precoding assume the number of RF chains $R = N$. The proposed method is employed for different values of $R$, and the analog precoders are chosen from a 64$\times$64 DFT codebook using the MWASO method. The figure also comprises the SER achieved by the CI-based hybrid precoding with the continuous-valued CPC method. 

The figure reveals that the CI-based hybrid precoding (even with $R=K$, and codebook-based analog precoders) yields significantly better performance than the conventional fully-digital precoding. As we increase the number of RF chains, the SER of the CI-based hybrid precoding gradually approaches that of the optimal CI-based fully-digital precoding. Moreover, we notice that the continuous-valued analog precoding (CPC) yields considerably better results than the codebook-based analog precoding (MWASO) due to a larger number of degrees of freedom at the cost of expensive full-resolution PSs.   


\subsection{Evaluation of block-level analog precoding techniques}
\label{subsec_results_block_level_analog_precoding}

\begin{figure}[b!]
	\centering
	\includegraphics[scale=1]{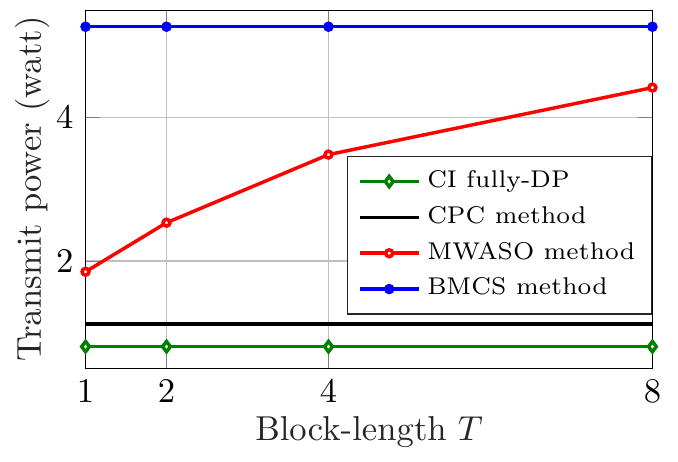}
	\caption{Comparison of different block-level analog precoding techniques for $N=64$, $R=K=16$, $M=4$, $\mathbf{E}=\mathbf{0}$, TNR = 1, and $T_{\text{c}}=8$.}
	\label{fig_P_vs_BL_vs_AP_methods}
\end{figure}

In this subsection, we compare the performance of different block-level analog precoding techniques. In order to facilitate a fair comparison, the optimal CI-based digital precoders are designed followed by each block-level analog precoding technique. Fig.~\ref{fig_P_vs_BL_vs_AP_methods} plots the transmit power of the hybrid precoders employing different analog precoding methods over a range of block-length $T$. In the simulation, we assume the channel is constant for $T_{\text{c}}=8$ symbol-intervals.  As we discussed in Section~\ref{sec_analog_precoders_design}, the CPC and BMCS methods are solely based on the channel matrix. Therefore, the transmit powers associated with these methods are constant over the block-length $T$. On the contrary, the MWASO method designs the analog precoders based on both the channel matrix and the transmit symbol vectors. Thus, the transmit power required by the MWASO method increases with the increase in $T$. In the figure, we notice that the continuous value-based CPC method outperforms the codebook-based MWASO and BMCS methods due to a larger number of degrees of freedom at the cost of expensive full-resolution PSs. Among the codebook-based methods the MWASO method, which exploits both channel and data information, outperforms the channel-only-based BMCS method. The MWASO method facilitates a performance-complexity trade-off based on the value of $T$. It also serves as a benchmark to quantify the performance-loss incurred by the low-complexity BMCS method. As $T$ increases the MWASO method needs to accommodate a large number of transmit symbol vectors, and hence its performance approaches that of the BMCS method. Therefore, the MWASO method is most appropriate when $T \ll T_c$.


\subsection{Computational complexity analysis}
\label{subsec_computtional_complexity}

In this paper, we derived closed-form expressions for the descent-direction, which can be computed in parallel, and the step-size, resulting in a single-layer iterative algorithm to solve problem (\ref{problem_update_base_precoder}), instead of a computationally more demanding two-layer iterative algorithm as in \cite{exploiting_masouros}. The computational complexity of Alg.~\ref{alg_iterative_parallel_method} (computing $\hat{\boldsymbol{\lambda}}$ in Eq.~(\ref{eq_closed_form_expression_lambda_hat}), $\eta^{\topindex{p}}$ in Eq.~(\ref{eq_step_size}), and  $\boldsymbol{\lambda}^{\topindex{p+1}}$ in Eq.~(\ref{eq_update_lambda})) is $\order{NW^2}$ for each iteration. The computation of $\mathbf{g}^\star$ and $\mathbf{N}$ involves the inversion of Hermitian matrix $\mathbf{M}_0^{\trans}\mathbf{M}_0$ of size $2R\times 2R$, incurring a complexity of $\order{R^3}$. We notice that the value of $R$ (no. of RF chains) is expected to be relatively small, and this inversion can be computed only once and reused in all subsequent iterations. In Alg.~\ref{alg_robust_algorithm}, the complexity associated with computing the worst-case error matrices and the corresponding optimal values for all users in each iteration (equations (\ref{eq_anti-clockwise_error_matrix}), (\ref{eq_clockwise_error_matrix}), (\ref{eq_vk_plus}), and (\ref{eq_vk_minus}) ) is $\order{NRK}$. 

In the following, we numerically evaluate the computational complexity of different methods discussed in this paper in terms of their computational time. The simulations are conducted on a system having the following features: Intel (R) Core (TM) i7-4790K CPU $@$ 4.00GHz, Arch Linux 4.16.8, MATLAB 2018b. 

\begin{table}[h!]
	\caption{The geometric mean of computational time (in milliseconds) to implement the proposed robust hybrid precoding using the interior-point method and the proposed scheme, for $N=128$, $R=K=M=4$, TNR = 1, $T=1$, and the CPC analog precoding method.}
	\begin{center}
		\begin{tabular}{| l || c | c | c | c | c | c |} \hline            
			Phase error bound $\delta$ & 0\textdegree  & 1\textdegree & 2\textdegree & 3\textdegree & 4\textdegree   \\ \hline \hline
			Interior-point-convex (\textit{quadprog}) & 2.44 & 4.52 & 4.50 & 4.51 & 4.55  \\ \hline
			Proposed scheme (Alg.~\ref{alg_iterative_parallel_method}) & 1.20 & 3.08 & 3.00 & 3.12 & 3.34   \\ \hline                        
		\end{tabular}
	\end{center}
	\label{table_robust_time_comparison}
\end{table}

Table~\ref{table_robust_time_comparison} lists the geometric mean of computational time required to implement the proposed robust hybrid precoding using the interior-point-convex (invoked from \textit{quadprog} function in MATLAB) and the proposed scheme (Alg.~\ref{alg_iterative_parallel_method}) for different $\delta$. In the table, we notice that the proposed scheme, which is customized to exploit the structure in the problem, is faster (saves an average of approx. 35\% of the computational time) than the general purpose interior-point method.


\section{Conclusion}
\label{sec_conclusion}

In this paper, we developed an algorithm for computing the optimal CI-based digital precoders with robustness against errors in the PSs. We also devised a low-complexity scheme to facilitate the implementation of the proposed algorithm efficiently in a distributed manner on parallel hardware architectures. Furthermore, we proposed block-level analog precoding techniques, which are necessary for ultra-low latency applications. The simulation results demonstrated the advantage of the proposed precoding over a conventional robust hybrid precoding method. The results illustrated the superiority of the CI-based hybrid precoding when compared to the interference suppression-based state-of-the-art schemes. We also verified that the devised scheme is faster in implementing the robust precoding when compared to a general purpose solver. 
Furthermore, we inferred from the simulations that the continuous-valued analog precoders yield significantly better performance, at the cost of high-resolution PSs, when compared to the codebook-based analog precoders. 


\appendices
\section{Closed-form expressions for the worst-case error matrices}
\label{appendix_closed_form_solutions}
Consider the objective function of problem (\ref{problem_anti-clockwise_error_matrix})
	\begin{subequations}
	\begin{flalign*}
		  \hat{f} \bydef &\operatorname{Im}\left(\mathbf{h}_k^{\trans} (\hat{\mathbf{A}} \odot \mathbf{E}) \mathbf{b}^{\topindex{i} \star} \right)  - \left(\operatorname{Re}\left(\mathbf{h}_k^{\trans} (\hat{\mathbf{A}} \odot \mathbf{E}) \mathbf{b}^{\topindex{i} \star} \right) -  \gamma_k\right) \tan\theta.
	\end{flalign*}
\end{subequations}
Let $g \bydef \mathbf{h}_k^{\trans} (\hat{\mathbf{A}} \odot \mathbf{E}) \mathbf{b}^{\topindex{i} \star}$. We can rewrite $g$ as, $g =   \sum_{\frall{n}} \sum_{\frall{r}} h_{kn} b_{r}^{\topindex{i}\star} \hat{a}_{nr} e_{nr}$, where $h_{kn}$ denotes the $n$th element of vector $\mathbf{h}_k$. It reveals that the objective function $\hat{f}$ is separable in each optimization variable $e_{nr}$. Therefore, $\hat{f}$ can be maximized separately and individually w.r.t. each $e_{nr}$ for $n \in \mathcal{N}, r \in \mathcal{R}$. Consider a summand $h_{kn} b_{r}^{\topindex{i}\star} \hat{a}_{nr} e_{nr}$ of $g$. Define $\bar{\chi} + j\tilde{\chi} \bydef h_{kn} b_{r}^{\topindex{i}\star} \hat{a}_{nr}$, and $\alpha + j\beta \bydef e_{nr} $. Substituting these new definitions, the part of function $\hat{f}$ that comprises the variable $e_{nr}$ can be expressed as 
\begin{align*}
\tilde{f}(\alpha, \beta) = \underbrace{(\tilde{\chi} - \bar{\chi}\tan\theta)}_{\kappa}  \alpha + &\underbrace{(\bar{\chi} + \tilde{\chi}\tan\theta)}_{\tau} \beta.
\end{align*}
The constraints on phase error values, given by $\abs{e_{nr}} = 1$ and $\abs{\angle e_{nr}} \leq \delta$, can be equivalently expressed as $\alpha^2 + \beta^2 = 1$ and $\alpha\geq \cos\delta$. Substituting $\beta = \pm \sqrt{1- \alpha^2}$ in the above equation, we get a new equivalent function $f(\alpha) = \kappa \alpha \pm \tau \sqrt{1-\alpha^2}$. This function comprises the following two variants: $
f_1(\alpha) = \kappa \alpha + \tau \sqrt{1-\alpha^2}$, and $f_2(\alpha) = \kappa \alpha - \tau \sqrt{1-\alpha^2}$. 
Note that $\sqrt{1-\alpha^2}$ is a concave function \cite{convex_boyd}. We can identify two cases based on the value of $\tau$. In the first case when $\tau \geq 0$, $f_1$ is a concave function, $f_2$ is a convex function, and $f_1 \geq f_2$ for $\abs{\alpha} \leq 1$. Moreover, an optimal point $\alpha^\star$ that maximizes $f_1$ also maximizes $\tilde{f}$ together with $\beta^\star = \sqrt{1-{\alpha^{\star}}^2}$. Similarly, we argue that in the second case when $\tau \leq 0$, an optimal point $\alpha^\star$ that maximizes (then) concave function $f_2$ also maximizes $\tilde{f}$ together with $\beta^\star = -\sqrt{1-{\alpha^\star}^2}$. 

If $\tau \geq 0$, we can obtain the optimal point $\alpha^\star$ that maximizes $f_1$, by differentiating $f_1$ w.r.t. $\alpha$ and equating to zero, i.e., $\frac{df_1}{d\alpha} = \kappa - \frac{\tau \alpha}{\sqrt{1-\alpha^2}} = 0 \implies \alpha^\star = \frac{\kappa}{\sqrt{\kappa^2 + \tau^2}}$.
Similarly, if $\tau \leq 0$ we can obtain the optimal point $\alpha^\star$ that maximizes $f_2$ as $\frac{df_2}{d\alpha} = \kappa + \frac{\tau \alpha}{\sqrt{1-\alpha^2}} = 0 \implies \alpha^\star = \frac{\kappa}{\sqrt{\kappa^2 + \tau^2}}$.
(In the previous derivations, we have explicitly used the prior knowledge of sign of $\tau$ and used the intermediate result that reveals the sign of $\alpha$ should be the same as the sign of $\kappa$).

Remember the function $f_1$ is concave in $\alpha$ if $\tau \geq 0 $ and $f_2$ is concave in $\alpha$ if $\tau \leq 0$. Therefore, if the obtained optimal point $\alpha^\star$ is smaller than $\cos\delta$ then we can enforce $\alpha^\star = \cos\delta$ to get the optimal point within the domain of the phase error that maximizes $\tilde{f}$. Substituting the expressions for $\kappa$ and $\tau$ we get $e_{nr}^{\star} = \alpha^\star + j\beta^\star$, where
\begin{align*}
&\alpha^\star = \operatorname{max}\left( \cos\delta, \ \ \dfrac{\tilde{\chi}\cos\theta - \bar{\chi} \sin\theta}{\abs{\bar{\chi} +j\tilde{\chi}}} \right), \\
&\beta^\star = \text{sign}(\tau) \sqrt{1- {\alpha^\star}^2} = \dfrac{\bar{\chi} + \tilde{\chi}\tan\theta}{\abs{\bar{\chi} + \tilde{\chi}\tan\theta}} \sqrt{1- {\alpha^\star}^2}.
\end{align*}

Let $\mathbf{Z} \bydef \left(\mathbf{h}_{k} {(\mathbf{b}^{\topindex{i} \star})}^{\trans}\right) \odot \hat{\mathbf{A}}$. Then, the worst-case error values for all PSs at the BS can be obtained efficiently by computing the error matrix $\mathbf{E}_{k}^{\plusindex{+}} = \mathbf{U}^{+} + j\mathbf{W}^{+}$, where
\begin{align*}
&{u}_{nr}^{+} = \operatorname{max}\left( \cos\delta, \ \ \dfrac{\operatorname{Im}({z}_{nr}) \cos\theta - \operatorname{Re}({z}_{nr})\sin\theta }{\abs{{z}_{nr}}} \right), \\
&{w}_{nr}^{+} = \dfrac{\operatorname{Re}({z}_{nr}) + \operatorname{Im}({z}_{nr}) \tan\theta}{\abs{\operatorname{Re}({z}_{nr}) + \operatorname{Im}({z}_{nr}) \tan\theta }} \sqrt{1 - ({u}_{nr}^{+})^2}.
\end{align*}
Similarly, we can derive the expression for $\mathbf{E}_{k}^{\minusindex{-}} = \mathbf{U}^{-} + j\mathbf{W}^{-}$, where
\begin{align*}
&{u}_{nr}^{-} = \operatorname{max}\left( \cos\delta, \ \ \dfrac{-\operatorname{Im}({z}_{nr}) \cos\theta - \operatorname{Re}({z}_{nr})\sin\theta }{\abs{{z}_{nr}}} \right), \\
&{w}_{nr}^{-} = \dfrac{-\operatorname{Re}({z}_{nr}) + \operatorname{Im}({z}_{nr}) \tan\theta}{\abs{-\operatorname{Re}({z}_{nr}) + \operatorname{Im}({z}_{nr}) \tan\theta } } \sqrt{1 - ({u}_{nr}^{-})^2}.
\end{align*}


\section{Convergence properties of Alg.~\ref{alg_robust_algorithm}}

\textbf{Lemma 1}: \textit{The problems (\ref{problem_robust_virtual_multicast_problem}) and (\ref{problem_update_base_precoder}) have unique global optimal solutions.} \\
Proof: The analog precoder matrix $\hat{\mathbf{A}}$ in problem (\ref{problem_robust_virtual_multicast_problem}) is a full column rank matrix. In case $\hat{\mathbf{A}}$ is not a full column rank matrix, it can be easily converted into a full column rank matrix without altering the effective hybrid precoder $\hat{\mathbf{A}}\mathbf{b}$ as follows:
Let $\hat{\mathbf{A}} \bydef [\hat{\mathbf{a}}_1, \ldots, \hat{\mathbf{a}}_R] \in \mathbb{C}^{N \times R}$, and a vector $\mathbf{b} \bydef [b_1, \ldots, b_R]^{\trans}$ where $N \geq R$. W.l.o.g. let the rank of $\hat{\mathbf{A}}$ be $R-1$, with $\hat{\mathbf{a}}_R$ being linearly dependent on other analog precoders in the matrix $\hat{\mathbf{A}}$, i.e., $\hat{\mathbf{a}}_R = w_1 \hat{\mathbf{a}}_1 + \ldots + w_{R-1} \hat{\mathbf{a}}_{R-1}$, where $w_1, \ldots w_{R-1}$ are scalars. Then we have\\
$\hat{\mathbf{A}}\mathbf{b} = b_1 \hat{\mathbf{a}}_1 + \ldots + b_R\hat{\mathbf{a}}_{R}$ \\
$\phantom{\hat{\mathbf{A}}\mathbf{b}} = b_1 \hat{\mathbf{a}}_1 + \ldots + b_R(w_1 \hat{\mathbf{a}}_1 + \ldots + w_{R-1} \hat{\mathbf{a}}_{R-1})$ \\
$\phantom{\hat{\mathbf{A}}\mathbf{b}} = (b_1 + b_Rw_1) \hat{\mathbf{a}}_1 + \ldots + (b_{R-1} + b_Rw_{R-1} )\hat{\mathbf{a}}_{R-1}$ \\
$\phantom{\hat{\mathbf{A}}\mathbf{b}} = b'_1 \hat{\mathbf{a}}_1 + \ldots + b'_{R-1} \hat{\mathbf{a}}_{R-1} = \hat{\mathbf{A}}'\mathbf{b}'$,\\ 
where $\hat{\mathbf{A}}' \in \mathbb{C}^{N \times (R-1)}$ is a full column rank matrix.

Due to the full rank property of $\hat{\mathbf{A}}$, the matrix $\hat{\mathbf{A}}^{\hermi}\hat{\mathbf{A}}$ is positive definite, and the quadratic objective function $||\hat{\mathbf{A}}\mathbf{b}||^2 = \mathbf{b}^{\hermi}\hat{\mathbf{A}}^{\hermi}\hat{\mathbf{A}}\mathbf{b}$ is a strictly convex function in $\mathbf{b}$. Therefore, the  problem (\ref{problem_robust_virtual_multicast_problem}) has a unique global optimal solution \cite{convex_boyd}. Following the same reasoning, we can prove that problem (\ref{problem_update_base_precoder}) also has a unique global optimal solution in the $i$th iteration.

\textbf{Lemma 2:} \textit{a) The optimal solution of problem (\ref{problem_robust_virtual_multicast_problem}) is a feasible solution of problem (\ref{problem_update_base_precoder}). 
b) The optimal value $P^{\star}$ of problem (\ref{problem_robust_virtual_multicast_problem}) is an upper-bound for problem (\ref{problem_update_base_precoder}). }\\
Proof: Let $\mathbf{b}^\star$ and $P^\star$ be the optimal solution and the optimal value of problem (\ref{problem_robust_virtual_multicast_problem}) respectively. 
In (\ref{problem_update_base_precoder}), the union of set of error matrices in the $i$th iteration $\boldsymbol{\tilde{\mathcal{E}}}^i \bydef \{\boldsymbol{\mathcal{E}}^{\plusindex{i+}}_1 \cup \boldsymbol{\mathcal{E}}^{\minusindex{i-}}_1 \cup \ldots \cup \boldsymbol{\mathcal{E}}^{\plusindex{i+}}_K \cup \boldsymbol{\mathcal{E}}^{\minusindex{i-}}_K \} \subset \boldsymbol{\mathcal{E}}$. Therefore, the set of constraints of (\ref{problem_update_base_precoder}) is a subset of the set constraints of  (\ref{problem_robust_virtual_multicast_problem}). Hence, the optimal solution $\mathbf{b}^\star$ of problem (\ref{problem_robust_virtual_multicast_problem}) is a feasible solution of problem (\ref{problem_update_base_precoder}). Moreover, the problems (\ref{problem_robust_virtual_multicast_problem}) and (\ref{problem_update_base_precoder}) have an identical objective function. Therefore, problem (\ref{problem_update_base_precoder}) is a relaxation of problem (\ref{problem_robust_virtual_multicast_problem}) and the optimal value of problem  (\ref{problem_robust_virtual_multicast_problem}) is an upper bound for problem (\ref{problem_update_base_precoder}).


\subsubsection{\textbf{Proof of Theorem 1}}
\label{appendix_proof_of_theorem_1}

Let $\mathbf{b}^{\topindex{I} \star}$ denote the optimal solution of problem (\ref{problem_update_base_precoder}) in the $I$th iteration. We assume that the constraints in (\ref{constraint_robust_virtual_multicast_problem_1}) and (\ref{constraint_robust_virtual_multicast_problem_2}) are fulfilled at $\mathbf{b} = \mathbf{b}^{\topindex{I} \star}$ for the worst-case error matrices $\mathbf{E}_{k}^{\plusindex{I+}}$ and $\mathbf{E}_{k}^{\minusindex{I-}}$, $\frall{k}$ (hence for all matrices in $\boldsymbol{\mathcal{E}}$), and the algorithm is terminated. Therefore, the optimal point $\mathbf{b}^{\topindex{I} \star}$ of problem (\ref{problem_update_base_precoder}) is a feasible point of problem (\ref{problem_robust_virtual_multicast_problem}). Since problem (\ref{problem_update_base_precoder}) is a relaxation (convex outer approximation) of problem (\ref{problem_robust_virtual_multicast_problem}), $\mathbf{b}^{\topindex{I} \star}$ is also an optimal solution for problem (\ref{problem_update_base_precoder}), and due to Lemma 1 we have $ \mathbf{b}^{\topindex{I} \star} = \mathbf{b}^\star$. 


\subsubsection{\textbf{Proof of Theorem 2}}
\label{appendix_proof_of_theorem_2}
Here we follow a similar line of arguments as in \cite{numerical_gustafson} to prove Theorem 2. If Alg.~\ref{alg_robust_algorithm} terminates after a finite number of $I$ iterations, then $\mathbf{b}^{\topindex{I}\star} = \mathbf{b}^{\star}$ according to Theorem 1, which confirms Theorem 2 in this case. On the other hand, if Alg.~\ref{alg_robust_algorithm} does not terminate after a finite number of iterations we want to prove that $\lim_{i\to\infty} f(\mathbf{b}^{\topindex{i}\star}) = P^{\star}$. Let  $\mathcal{B} \triangleq \{\mathbf{b}^{\topindex{1}\star}, \mathbf{b}^{\topindex{2}\star},\ldots\}$ be the infinite sequence of optimal points of problem (\ref{problem_update_base_precoder}). Since the problems (\ref{problem_robust_virtual_multicast_problem}) and (\ref{problem_update_base_precoder}) are feasible, the elements of $\mathcal{B}$ are bounded. Due to the practical power budget constraints we can argue w.l.o.g. that the elements of $\mathcal{B}$ are confined to a compact set. Therefore, the sequence $\mathcal{B}$ has limit points \cite{numerical_gustafson,an_iterative_wu}. Let $\hat{\mathbf{b}}$ be a limit point. Let $\boldsymbol{\hat{\mathcal{E}}}^{\plusindex{+}}_k$ be a set of error matrices that are associated with the constraint (\ref{constraint_update_base_precoder_1}) at point $\mathbf{b} = \hat{\mathbf{b}}$. For the purpose of contradiction assume $f(\hat{\mathbf{b}}) < P^{\star}$, i.e., $\hat{\mathbf{b}}$ is not a feasible point of problem (\ref{problem_robust_virtual_multicast_problem}). W.l.o.g. let $\bar{\mathbf{E}}^{\plusindex{+}}_k$ be a worst-case error matrix of the $k$th user that violates the constraint (\ref{constraint_robust_virtual_multicast_problem_1}) at point $\mathbf{b} = \hat{\mathbf{b}}$. Define the function associated with the constraint (\ref{constraint_robust_virtual_multicast_problem_1}) as 
$\hat{f}(\mathbf{b}, \mathbf{E}) \triangleq \operatorname{Im}\left(\mathbf{h}_k^{\trans} (\hat{\mathbf{A}} \odot \mathbf{E}) \mathbf{b} \right) - \left(\operatorname{Re}\left(\mathbf{h}_k^{\trans} (\hat{\mathbf{A}} \odot \mathbf{E}) \mathbf{b} \right) -  \gamma_k\right) \tan\theta$. Therefore, we have $\hat{f}(\hat{\mathbf{b}}, \bar{\mathbf{E}}^{\plusindex{+}}_k) > 0$. Moreover, we have $\hat{f}(\hat{\mathbf{b}}, \mathbf{E}) \leq 0, \forall \mathbf{E} \in \boldsymbol{\hat{\mathcal{E}}}^{\plusindex{+}}_k$. Consider a point $\mathbf{b}^{\topindex{i}\star} \in \mathcal{B}$ with a worst-case error matrix $\mathbf{E}_{k}^{\plusindex{i+}}$, and a subsequence $\mathbf{b}^{\topindex{i}\star} \rightarrow \hat{\mathbf{b}}$ in $\mathcal{B}$. Since $\boldsymbol{\mathcal{E}}$ is a compact set, we have a corresponding subsequence of worst-case error matrices  $\mathbf{E}_{k}^{\plusindex{i+}} \rightarrow \hat{\mathbf{E}}$ in set $\boldsymbol{\hat{\mathcal{E}}}^{\plusindex{+}}_k$ \cite{numerical_gustafson, an_iterative_wu}. By definition of $\mathbf{E}_{k}^{\plusindex{i+}}$ we have $\hat{f}(\mathbf{b}^{\topindex{i}\star}, \bar{\mathbf{E}}^{\plusindex{+}}_k)  \leq \hat{f}(\mathbf{b}^{\topindex{i}\star}, \mathbf{E}_{k}^{\plusindex{i+}})$. Letting $i \rightarrow \infty$ we get $\hat{f}(\hat{\mathbf{b}}, \bar{\mathbf{E}}^{\plusindex{+}}_k)  \leq \hat{f}(\hat{\mathbf{b}}, \hat{\mathbf{E}})$. It results in a contradicting result $0 < \hat{f}(\hat{\mathbf{b}}, \bar{\mathbf{E}}^{\plusindex{+}}_k)  \leq \hat{f}(\hat{\mathbf{b}}, \hat{\mathbf{E}}) \leq 0$. Hence, $f(\hat{\mathbf{b}}) < P^{\star}$ is not possible. Moreover, due to Lemma 2b we have $\lim_{i\to\infty} f(\mathbf{b}^{\topindex{i}\star}) = P^{\star}$.


\section{}
\label{appendix_ser_snr_tnr_relation}

For obtaining the relation between SER and SNR in the SNR/SINR fulfillment-based precoding system, Rayleigh-fading-based complex channels with zero mean and unit-variance are assumed. The unit-norm transmit symbols are drawn from the corresponding constellation set.
The i.i.d. complex Gaussian noise with zero mean and an appropriate variance are added to the received signal. The channel inversion and projection methods are employed to estimate the transmit symbols \cite{fundamental_tse}.
To obtain the relation between SER and TNR in a CI-based precoding system, the received signals are randomly generated on the threshold-margin (set to 1) of all symbols and i.i.d. complex Gaussian noise with zero mean and an appropriate variance are added to them.      
\begin{figure}[h!]
	\centering
	\includegraphics[scale=0.8]{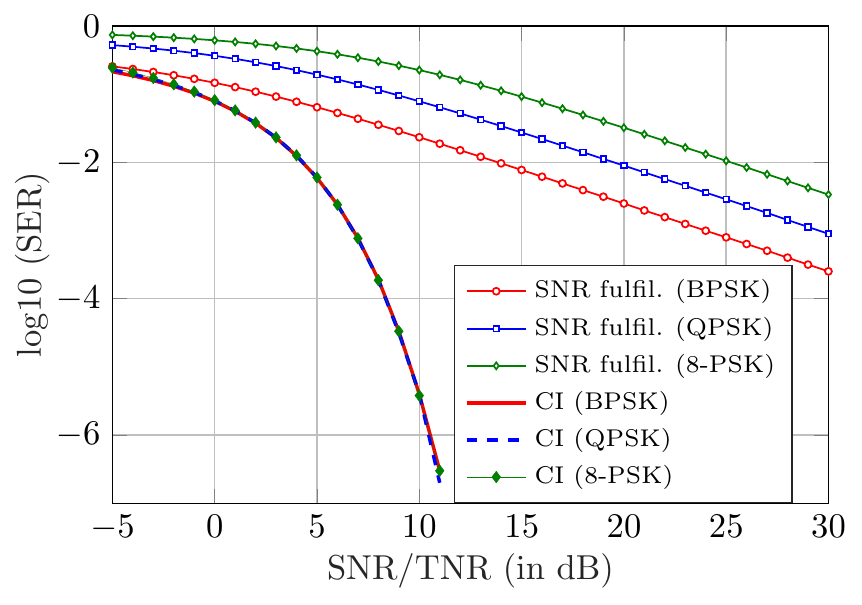}
	\caption{Empirical relation between SNR, TNR, and SER.}
	\label{fig_ser_snr_tnr}
\end{figure}




\section*{Acknowledgment}
The authors would like to thank Minh Trinh Hoang for valuable discussions related to the convergence proof of the algorithm. 


\bibliographystyle{IEEEtran}
\FloatBarrier
{\small \bibliography{ms}}

\begin{thebibliography}{10}
\providecommand{\url}[1]{#1}
\csname url@samestyle\endcsname
\providecommand{\newblock}{\relax}
\providecommand{\bibinfo}[2]{#2}
\providecommand{\BIBentrySTDinterwordspacing}{\spaceskip=0pt\relax}
\providecommand{\BIBentryALTinterwordstretchfactor}{4}
\providecommand{\BIBentryALTinterwordspacing}{\spaceskip=\fontdimen2\font plus
\BIBentryALTinterwordstretchfactor\fontdimen3\font minus
  \fontdimen4\font\relax}
\providecommand{\BIBforeignlanguage}[2]{{%
\expandafter\ifx\csname l@#1\endcsname\relax
\typeout{** WARNING: IEEEtran.bst: No hyphenation pattern has been}%
\typeout{** loaded for the language `#1'. Using the pattern for}%
\typeout{** the default language instead.}%
\else
\language=\csname l@#1\endcsname
\fi
#2}}
\providecommand{\BIBdecl}{\relax}
\BIBdecl

\bibitem{massive_mimo_in_hoydis}
J.~Hoydis, S.~ten Brink, and M.~Debbah, ``Massive {MIMO} in the {UL/DL} of
  cellular networks: {H}ow many antennas do we need?'' \emph{{IEEE} J. Select.
  Areas Commun.}, vol.~31, no.~2, pp. 160--171, Feb. 2013.

\bibitem{an_overview_lu}
L.~Lu \emph{et~al.}, ``An overview of massive {MIMO}: {B}enefits and
  challenges,'' \emph{{IEEE} J. Select. Topics in Signal Process.}, vol.~8,
  no.~5, pp. 742--758, Oct. 2014.

\bibitem{multiple_mietzner}
J.~Mietzner \emph{et~al.}, ``Multiple-antenna techniques for wireless
  communications---{A} comprehensive literature survey,'' \emph{IEEE Commun.
  Surveys Tutorials}, vol.~11, no.~2, pp. 87--105, Feb. 2009.

\bibitem{optimal_bengtsson}
M.~Bengtsson and B.~Ottersten, ``Optimal and suboptimal transmit beamforming,''
  \emph{Handbook of Antennas in Wireless Commun.}, Aug. 2001.

\bibitem{design_doan}
C.~H. Doan \emph{et~al.}, ``Design considerations for 60 {GHz} {CMOS} radios,''
  \emph{{IEEE} Commun. Mag.}, vol.~42, no.~12, pp. 132--140, Dec. 2004.

\bibitem{low_complexity_liang}
L.~Liang, W.~Xu, and X.~Dong, ``Low-complexity hybrid precoding in massive
  multiuser {MIMO} systems,'' \emph{{IEEE} Wireless Commun. Letters}, vol.~3,
  no.~6, pp. 653--656, Dec. 2014.

\bibitem{hybrid_digital_sohrabi}
F.~Sohrabi and W.~Yu, ``Hybrid digital and analog beamforming design for
  large-scale antenna arrays,'' \emph{{IEEE} J. Select. Topics in Signal
  Process.}, vol.~10, no.~3, pp. 501--513, Apr. 2016.

\bibitem{variable_zhang}
X.~Zhang, A.~F. Molisch, and S.~Y. Kung, ``Variable-phase-shift-based
  {RF}-baseband codesign for {MIMO} antenna selection,'' \emph{{IEEE} Trans.
  Signal Process.}, vol.~53, no.~11, pp. 4091--4103, Nov. 2005.

\bibitem{joint_li}
Z.~Li \emph{et~al.}, ``Joint optimization of hybrid beamforming for multi-user
  massive {MIMO} downlink,'' \emph{{IEEE} Trans. Wireless Commun.}, vol.~17,
  no.~6, pp. 3600--3614, Jun. 2018.

\bibitem{hybrid_molisch}
A.~F. Molisch \emph{et~al.}, ``Hybrid beamforming for massive {MIMO}: {A}
  survey,'' \emph{{IEEE} Commun. Mag.}, vol.~55, no.~9, pp. 134--141, Sep.
  2017.

\bibitem{hybrid_beamforming_hegde}
G.~Hegde, Y.~Cheng, and M.~Pesavento, ``Hybrid beamforming for large-scale
  {MIMO} systems using uplink-downlink duality,'' in \emph{Proc. {IEEE} Int.
  Conf. on Acoustics, Speech and Signal Process. {(ICASSP)}}, New Orleans, USA,
  Mar. 2017.

\bibitem{hybrid_analog_sohrabi}
F.~Sohrabi and W.~Yu, ``Hybrid analog and digital beamforming for {mmWave}
  {OFDM} large-scale antenna arrays,'' \emph{{IEEE} J. Select. Areas Commun.},
  vol.~35, no.~7, pp. 1432--1443, Jul. 2017.

\bibitem{exploiting_masouros}
C.~Masouros and G.~Zheng, ``Exploiting known interference as green signal power
  for downlink beamforming optimization,'' \emph{{IEEE} Trans. Signal
  Process.}, vol.~63, no.~14, pp. 3628--3640, Jul. 2015.

\bibitem{known_masouros}
C.~Masouros \emph{et~al.}, ``Known interference in the cellular downlink: {A}
  performance limiting factor or a source of green signal power?'' \emph{{IEEE}
  Commun. Mag.}, vol.~51, no.~10, pp. 162--171, Oct. 2013.

\bibitem{large_scale_amadori}
P.~V. Amadori and C.~Masouros, ``Large scale antenna selection and precoding
  for interference exploitation,'' \emph{{IEEE} Trans. Commun.}, vol.~65,
  no.~10, pp. 4529--4542, Oct. 2017.

\bibitem{rethinking_zheng}
G.~Zheng \emph{et~al.}, ``Rethinking the role of interference in wireless
  networks,'' \emph{{IEEE} Commun. Mag.}, vol.~52, no.~11, pp. 152--158, Nov.
  2014.

\bibitem{exploiting_timotheou}
S.~Timotheou, G.~Zheng, C.~Masouros, and I.~Krikidis, ``Exploiting constructive
  interference for simultaneous wireless information and power transfer in
  multiuser downlink systems,'' \emph{{IEEE} J. Select. Areas Commun.},
  vol.~34, no.~5, pp. 1772--1784, May 2016.

\bibitem{solution_schubert}
M.~Schubert and H.~Boche, ``Solution of the multiuser downlink beamforming
  problem with individual {SINR} constraints,'' \emph{{IEEE} Trans. Veh.
  Technol.}, vol.~53, no.~1, pp. 18--28, Jan. 2004.

\bibitem{analog_hegde}
G.~Hegde, C.~Masouros, and M.~Pesavento, ``Analog beamformer design for
  interference exploitation based hybrid beamforming,'' in \emph{Proc. {IEEE}
  Sensor Array and Multi. Signal Process. Workshop {(SAM)}}, Sheffield, UK,
  Jul. 2018.

\bibitem{towards_vannithamby}
R.~Vannithamby and S.~Talwar, \emph{Towards {5G}: {A}pplications, Requirements
  \& Candidate Technologies}.\hskip 1em plus 0.5em minus 0.4em\relax John Wiley
  \& Sons Incorporated, 2017.

\bibitem{robust_wu}
S.~Wu \emph{et~al.}, ``Robust hybrid beamforming with phased antenna arrays for
  downlink {SDMA} in indoor 60 {GHz} channels,'' \emph{{IEEE} Trans. Wireless
  Commun.}, vol.~12, no.~9, pp. 4542--4557, Sep. 2013.

\bibitem{constructive_khandaker}
M.~R.~A. Khandaker, C.~Masouros, and K.~K. Wong, ``Constructive interference
  based secure precoding: {A} new dimension in physical layer security,''
  \emph{{IEEE} Trans. on Inform. Forensics and Security}, vol.~13, no.~9, pp.
  2256--2268, Sep. 2018.

\bibitem{nonlinear_bertsekas}
D.~P. Bertsekas, \emph{Nonlinear programming}, 2nd~ed.\hskip 1em plus 0.5em
  minus 0.4em\relax Athena scientific Belmont, 1999.

\bibitem{exploiting_li}
A.~Li and C.~Masouros, ``Exploiting constructive mutual coupling in {P2P}
  {MIMO} by analog-digital phase alignment,'' \emph{{IEEE} Trans. Wireless
  Commun.}, vol.~16, no.~3, pp. 1948--1962, Mar. 2017.

\bibitem{sources_morton}
M.~A. Morton \emph{et~al.}, ``Sources of phase error and design considerations
  for silicon-based monolithic high-pass/low-pass microwave phase shifters,''
  \emph{{IEEE} Trans. on Microwave Theory and Techniques}, vol.~54, no.~12, pp.
  4032--4040, Dec. 2006.

\bibitem{60_GHz_li}
W.~Li \emph{et~al.}, ``60-{GHz} 5-bit phase shifter with integrated {VGA}
  phase-error compensation,'' \emph{{IEEE} Trans. on Microwave Theory and
  Techniques}, vol.~61, no.~3, Mar. 2013.

\bibitem{scaling_rusek}
F.~Rusek \emph{et~al.}, ``Scaling up {MIMO}: {O}pportunities and challenges
  with very large arrays,'' \emph{{IEEE} Signal Process. Mag.}, vol.~30, no.~1,
  pp. 40--60, Jan. 2013.

\bibitem{a_coordinated_yin}
H.~Yin, D.~Gesbert, M.~Filippou, and Y.~Liu, ``A coordinated approach to
  channel estimation in large-scale multiple-antenna systems,'' \emph{{IEEE} J.
  Select. Areas Commun.}, vol.~31, no.~2, pp. 264--273, Feb. 2013.

\bibitem{dynamic_masouros}
C.~Masouros and E.~Alsusa, ``Dynamic linear precoding for the exploitation of
  known interference in {MIMO} broadcast systems,'' \emph{{IEEE} Trans.
  Wireless Commun.}, vol.~8, no.~3, pp. 1396--1404, Mar. 2009.

\bibitem{generalized_vazquez}
F.~G. Vázquez, J.-J. Rückmann, O.~Stein, and G.~Still, ``Generalized
  semi-infinite programming: {A} tutorial,'' \emph{Journal of Computational and
  Applied Mathematics}, vol. 217, no.~2, pp. 394--419, 2008.

\bibitem{numerical_gustafson}
S.~A. Gustafson and K.~O. Kortanek, ``Numerical treatment of a class of
  semi-infinite programming problems,'' \emph{Naval Research Logistics
  Quarterly}, vol.~20, no.~3, pp. 477--504, 1973.

\bibitem{relaxed_wu}
S.~Y. Wu, S.~C. Fang, and C.~J. Lin, ``Relaxed cutting plane method for solving
  linear semi-infinite programming problems,'' \emph{Journal of Optimization
  Theory and Applications}, vol.~99, no.~3, pp. 759--779, 1998.

\bibitem{quality_karipidis}
E.~Karipidis, N.~D. Sidiropoulos, and Z.~Q. Luo, ``Quality of service and
  {Max-Min} fair transmit beamforming to multiple cochannel multicast groups,''
  \emph{{IEEE} Trans. Signal Process.}, vol.~56, no.~3, pp. 1268--1279, Mar.
  2008.

\bibitem{aunified_yang}
Y.~Yang and M.~Pesavento, ``A unified successive pseudoconvex approximation
  framework,'' \emph{{IEEE} Trans. Signal Process.}, vol.~65, no.~13, pp.
  3313--3328, Jul. 2017.

\bibitem{parallel_hegde}
G.~Hegde, Y.~Yang, C.~Steffens, and M.~Pesavento, ``Parallel low-complexity
  {M-PSK} detector for large-scale {MIMO} systems,'' in \emph{Proc. {IEEE}
  Sensor Array and Multi. Signal Process. Workshop {(SAM)}}, Rio de Janeiro,
  Brazil, Jul. 2016, pp. 1--5.

\bibitem{convex_boyd}
S.~Boyd and L.~Vandenberghe, \emph{Convex Optimization}.\hskip 1em plus 0.5em
  minus 0.4em\relax New York, USA: Cambridge University Press, 2004.

\bibitem{atutorial_palomar}
D.~P. Palomar and M.~Chiang, ``A tutorial on decomposition methods for network
  utility maximization,'' \emph{{IEEE} J. Select. Areas Commun.}, vol.~24,
  no.~8, pp. 1439--1451, Aug. 2006.

\bibitem{limited_alkhateeb}
A.~Alkhateeb, G.~Leus, and R.~W. Heath, ``Limited feedback hybrid precoding for
  multi-user millimeter wave systems,'' \emph{{IEEE} Trans. Wireless Commun.},
  vol.~14, no.~11, pp. 6481--6494, Nov. 2015.

\bibitem{joint_hegde}
G.~Hegde and M.~Pesavento, ``Joint user selection and hybrid analog-digital
  beamforming in massive {MIMO} systems,'' in \emph{Proc. {IEEE} Sensor Array
  and Multi. Signal Process. Workshop {(SAM)}}, Sheffield, UK, Jul. 2018.

\bibitem{design_bansal}
D.~Bansal \emph{et~al.}, ``Design of novel compact anti-stiction and low
  insertion loss {RF MEMS} switch,'' \emph{Microsystem Technologies}, vol.~20,
  no.~2, pp. 337--340, Feb. 2014.

\bibitem{block_elhamifar}
E.~Elhamifar and R.~Vidal, ``Block-sparse recovery via convex optimization,''
  \emph{{IEEE} Trans. Signal Process.}, vol.~60, no.~8, pp. 4094--4107, Aug.
  2012.

\bibitem{block_steffens}
C.~Steffens and M.~Pesavento, ``Block- and rank-sparse recovery for direction
  finding in partly calibrated arrays,'' \emph{{IEEE} Trans. Signal Process.},
  vol.~66, no.~2, pp. 384--399, Jan. 2018.

\bibitem{hybrid_mimo_rial}
R.~Méndez-Rial \emph{et~al.}, ``Hybrid {MIMO} architectures for millimeter
  wave communications: {P}hase shifters or switches?'' \emph{IEEE Access},
  vol.~4, pp. 247--267, Jan. 2016.

\bibitem{spatially_ayach}
O.~E. Ayach \emph{et~al.}, ``Spatially sparse precoding in millimeter wave
  {MIMO} systems,'' \emph{{IEEE} Trans. Wireless Commun.}, vol.~13, no.~3, pp.
  1499--1513, Mar. 2014.

\bibitem{worst_case_wang}
J.~Wang and D.~P. Palomar, ``Worst-case robust {MIMO} transmission with
  imperfect channel knowledge,'' \emph{{IEEE} Trans. Signal Process.}, vol.~57,
  no.~8, pp. 3086--3100, Aug. 2009.

\bibitem{a_robust_pascual}
A.~Pascual-Iserte, D.~P. Palomar, A.~I. Perez-Neira, and M.~A. Lagunas, ``A
  robust maximin approach for {MIMO} communications with imperfect channel
  state information based on convex optimization,'' \emph{{IEEE} Trans. Signal
  Process.}, vol.~54, no.~1, pp. 346--360, Jan. 2006.

\bibitem{an_iterative_wu}
S.~Wu, D.~Li, L.~Qi, and G.~Zhou, ``An iterative method for solving {KKT}
  system of the semi-infinite programming,'' \emph{Optimization Methods and
  Software}, vol.~20, no.~6, pp. 629--643, 2005.

\bibitem{fundamental_tse}
D.~Tse and P.~Viswanath, \emph{Fundamentals of Wireless Communication}.\hskip
  1em plus 0.5em minus 0.4em\relax Cambridge University Press, 2005.

\end{thebibliography}


\end{document}